\documentclass[trackchanges]{aastex701}

\newcommand{\aox}{\ensuremath{\alpha_{\textsc{ox}}}}

\newcommand{\eddfrac}{\lambda_{\rm Edd}}
\newcommand{\aoxone}{\ensuremath{\alpha_{\textsc{ox}}^{(0.5)}}}
\newcommand{\aoxtwo}{\ensuremath{\alpha_{\textsc{ox}}^{(4)}}}

\usepackage{soul}

\usepackage{multirow}
\usepackage{amsmath}
\usepackage{longtable}

\begin{document}

\title{A break in the X-ray loudness of Markarian 590: evidence for an AGN spectral state transition?}

\author[orcid=0000-0002-4533-3170]{Biswaraj Palit}
\affiliation{Nicolaus Copernicus Astronomical Center, Polish Academy of Sciences, Bartycka 18, 00-716, Warsaw, Poland}
\email{bpalit@camk.edu.pl}  

\author[orcid=0000-0002-5275-4096]{Agata Różańska} 
\affiliation{Nicolaus Copernicus Astronomical Center, Polish Academy of Sciences, Bartycka 18, 00-716, Warsaw, Poland}
\email{agata@camk.edu.pl}

\author[orcid=0000-0002-2173-0673]{Alex G. Markowitz}
\affiliation{Nicolaus Copernicus Astronomical Center, Polish Academy of Sciences, Bartycka 18, 00-716, Warsaw, Poland}
\email{almarkowitz@camk.edu.pl}

\author[orcid=0000-0003-3959-5534]{Daniel Lawther}
\affiliation{DARK, Niels Bohr Institute, University of Copenhagen, Jagtvej 155, 2200, Copenhagen, Denmark}
\email{unclellama@gmail.com}

\author[0000-0001-9191-9837]{Marianne Vestergaard}
\affiliation{DARK, Niels Bohr Institute, University of Copenhagen, Jagtvej 155, 2200, Copenhagen, Denmark}
\affiliation{Steward Observatory and Department of Astronomy, University of Arizona, 933 N.\ Cherry Avenue, 85721, Tucson, AZ, USA}
\email{mvester@nbi.ku.dk}

\author[0000-0001-8665-5523]{John J. Ruan}
\affiliation{Bishop's University, Department of Physics \& Astronomy, 2600 rue College, Sherbrooke, J1M 1Z7, Québec, Canada}
\email{jruan@ubishops.ca}

\author[0000-0001-5647-3366]{Tathagata Saha}
\affiliation{Inter-University Centre for Astronomy and Astrophysics, SPPU Campus, 411007, Pune, India}
\email{tathagata@camk.edu.pl}

\author[0000-0003-1551-1340]{Gregory Walsh}
\affiliation{DARK, Niels Bohr Institute, University of Copenhagen, Jagtvej 155, 2200, Copenhagen, Denmark}
\email{gregory.walsh@nbi.ku.dk}

\author[0000-0002-9807-4520]{Abhijeet Borkar}
\affiliation{Astronomical Institute of the Czech Academy of Sciences, Bo\v{c}n\'i II 1401, CZ-14100 Prague, Czechia}
\email{abhijeet.borkar@asu.cas.cz}

\author[0000-0003-2656-6726]{Marzena \'Sniegowska}
\affiliation{Astronomical Institute of the Czech Academy of Sciences, Bo\v{c}n\'i II 1401, CZ-14100 Prague, Czechia}
\email{msniegowska@camk.edu.pl}

\author[0000-0002-2310-0982]{Kai-Xing Lu}
\affiliation{Yunnan Observatories, Chinese Academy of Sciences, Kunming, 650011, People’s Republic of China}
\email{lukx@ynao.ac.cn}

\correspondingauthor{Biswaraj Palit}
\email{bpalit@camk.edu.pl}

\begin{abstract}
Using decade-long multi-band {\it Swift} observations of the changing-look AGN Markarian 590, we identify a clear break in the dependence of the X-ray loudness parameter $\aox$ on the source accretion rate. This break could be potential evidence for an accretion state transition, analogous to that observed for X-ray binaries.
The $\aox$ follows a pronounced \lq V\rq-shaped dependence on Eddington ratio $\eddfrac$, with a statistically significant break at $\eddfrac =0.021\pm0.008$, consistent with the Eddington-ratio threshold associated with changing-look events in quasars. This behavior is indicative of a change in the inner accretion flow, from a truncated disk with a dominant hot corona at low accretion rates to an inward extending disk with enhanced UV emission and a prominent warm Comptonizing layer at higher rates. The UV and X-ray Eddington ratio tracers also show consistent breaks at $\eddfrac\sim0.004$. Mkn~590 evolves through distinct phenomenological accretion phases, from a faint, hard X-ray dominated state, through a flaring phase, to a bright, UV/soft X-ray dominated phase and exhibiting variability on month-, year-, and decade-long timescales. This overall evolution is shorter than classical viscous timescales but broadly consistent with propagating thermal fronts in the accretion disk. We also found a declining radio-to-X-ray luminosity ratio with increasing $\eddfrac$, indicating a relative suppression of radio emission as the disk becomes more dominant with respect to the X-ray corona. Taken together, these results provide evidence that Mkn~590 is undergoing a state transition, supporting a broad analogy between changing-look AGNs and X-ray binaries. \end{abstract}


\section{Introduction}
\label{sec:intro}
 One of the most compelling issues in accretion of matter onto compact objects is whether the nature of the accretion flow is universal across black holes spanning more than eight orders of magnitude in mass, from stellar mass black holes ($\sim$5--20 M$_{\odot}$) in X-Ray binaries (XRBs), through intermediate mass black holes ($\sim$ 10$^{3-5}$ M$_{\odot}$) to supermassive black holes ($\sim$10$^{6-10}$ M$_{\odot}$) at the center of active galactic nuclei (AGNs). Despite this enormous dynamic range, the broadband spectral energy distribution (SED) 
 frequently shows multiple emission components that are common across different object classes.
A common component
 of the SED of accreting black holes 
 is typically attributed to the thermal emission from an optically-thick, standard accretion disk,
 well described by a multi-color blackbody spectrum, where the peak flux  
is inversely proportional to the black hole mass \citep{1973ss}. In XRBs, a relatively higher disk temperature ($\sim 10^{7}$~K) causes the thermal emission to peak 
in the soft X-ray regime ($\leq 1$~keV). In contrast, SMBHs have cooler disks ($\sim10^{5}$~K), leading the thermal emission to peak in the extreme ultraviolet (EUV) band. 

The second common SED component occurs in hard X-rays, consisting of a high-energy tail 
 and it is commonly associated with an inverse Compton scattering of lower energy photons by a hot ($10^{8-9}$~K), optically-thin 
  corona \citep{1991haardt}.
A third, distinct spectral feature frequently observed in approximately 50\% of AGNs is the soft X-ray excess (hereafter SXE), which appears at energies below 2\,keV \citep{1997laor}. 
This component is typically attributed to the 
Comptonised emission from a warm corona \citep{2015rozanska} situated at the intersection of an
accretion disk and a hot corona.  Observations support the presence of a warm, optically thick corona \citep{POP2018A&A...611A..59P} which dominates the emission in the far-UV to soft X-ray band. Thus, in terms of energetics, this component may be interpreted as an extension or modification of the thermal disk emission. 
This effect is particularly relevant in  
the disks of XRBs, where the SXE overlaps spectrally with the thermal disk emission. However,  recently it was demonstrated that the presence of such a warm Comptonizing layer in XRBs resulted in systematically lower black hole's spin values \citep{2024zdziarski}, bringing them into closer agreement with spin estimates from black hole merger events by gravitational waves \citep{GW2023PhRvX..13a1048A}.
If the overall SED shape of accreting compact objects can indeed be explained by 
the three components (disk, hot and warm corona), then as the accretion regime changes with time, one would expect a consistent pattern of variability across XRBs to AGNs, reflecting a complicated interplay between the disk, the hot corona, and the warm corona.


XRBs are well known to undergo spectral state transitions. Generally, at low luminosities, sources are in a low/hard state, with X-ray spectra dominated by emission from the hot corona. However, during state transitions, rapid outbursts occur, with X-ray luminosity increasing by several orders of magnitude for weeks to months, and with sources frequently reaching a high/soft state, dominated by thermal disk emission (for reviews see \citet{Remillard2006ARA&A..44...49R, 2007Done}; see also \citealt{Esin1997ApJ...489..865E, poutanen1997MNRAS.292L..21P}). As the flares decay and mass accretion rate drops again, the inner flow collapses into an optically-thin, hot flow, probably through evaporation \citep{Rozanska2000A&A...360.1170R}. 
The evolution of the emission between these states can be studied using hardness-intensity diagrams, which represents the variation of the X-ray spectral hardness against the X-ray luminosity. These diagrams trace a characteristic \lq q\rq-shaped track over timescales of months--years and the distinct regions on this track signify separate spectral states \citep{homan2005Ap&SS.300..107H}. Attempts to identify analogous state transitions in AGNs have been challenging, primarily due to their much longer accretion timescales. 
So far, only weak indications of the \lq q\rq-shaped track were recovered for large samples studies involving AGNs \citep{2006kording,Ontiveros2021MNRAS.504.5726F,Svoboda2017A&A...603A.127S,Emily2022A&A...662A..28M}. 
 In this paper, we study the recently reactivated changing look AGN Markarian\,590 (hereafter Mkn\,590; \citealt{2025Lawther,2025Palit}). As we demonstrate here, it exhibits clear evidence of a spectral state transition, analogous to those observed in XRBs.

The X-ray loudness parameter, $\aox$, has been introduced as a proxy for comparing the coronal X-rays against the thermal disk UV luminosities in quasars  \citep{zamorani1981ApJ...245..357Z,1979tananbaum,Young2010ApJ...708.1388Y} and bright AGNs \citep{lusso2010A&A...512A..34L,Grupe2010ApJS..187...64G}. 
Simulated values of $\aox$ for SMBHs, based on canonical XRB spectral states, show a characteristic \lq V\rq-shaped dependence on the Eddington ratio  \citep{Sobo2011MNRAS.413.2259S}. Subsequent studies of quasars observed in bright and faint phases find broad consistency with this predicted trend \citep{Ruan2019ApJ...883...76R}. A similar evolution of $\aox$ with Eddington ratio is 
also reported for the tidal disruption event AT2018fyk, which showed an XRB-like spectral state transition during its outburst 
\citep{wevers2021ApJ...912..151W}.
In this paper, using optical, UV and X-ray data of Mkn~590 collected over twelve years \citep{2025Lawther} by the \textit{Neil Gehrels Swift Observatory}, we aim to determine the evolution of $\aox$ with Eddington ratio. Earlier, such a result was obtained only for the case of the CLAGN Mkn~1018 \citep{lyu2021MNRAS.506.4188L}, albeit with sparsely sampled data.
Other notable individual cases such as NGC 1566 \citep{2021MNRASjana}, Mkn 1018 \citep{saha2025A&A...699A.205S}, NGC~2617
and ZTF18aajupnt \citep{2019arXivRuan} provide preliminary evidence of a \lq V\rq-shaped $\aox$ trend, but comprehensive observational confirmation of a state transition in individual AGNs are still lacking. 


CLAGNs undergo dramatic changes in luminosity on  human observational timescales \citep[][and references therein]{2023Ricci}. These luminosity variations are accompanied by the appearance/disappearance of the broad emission lines in the optical regime, originating from the broad line region (BLR) \citep[first reported for Mkn~6 by][]{1971khachikian}. To date, nearly 1000 CLAGNs have been identified, exhibiting transitions from Seyfert Types 2, 1.9, 1.8, and 1.5 \citep[AGN optical spectral subtypes defined by][]{1977osterbrock,1981osterbrock} to Types 1 and 1.2 on timescales ranging from a few months to decades or longer \citep{2024Panda,DESIGuo_2025}. Thus, the study of these transient AGN is crucial for understanding the physics of accretion onto SMBHs, as the changing look event often challenges the standard inclination-based definition of the unified model of Seyfert galaxies \citep{anotnoucci1993ARA&A..31..473A}. 
In the majority of CLAGN, 
spectral and luminosity changes are believed to be driven by intrinsic variations in the accretion flow rate onto the SMBH \citep{Noda2018MNRAS.480.3898N,Ricci2020ApJ...898L...1R,2025Palit},
resulting in structural changes of the inner accretion flow and/or 
major variations in the net ionizing radiation that illuminates the BLR \citep{Noda2018MNRAS.480.3898N}.
It should be noted that the evolution of the BLR structure as a function of luminosity is suspected to be independent of changing look events \citep[e.g.][]{2009Elitzur}, and it is important for the community to investigate this issue in parallel with our present work. The required detailed spectroscopic analysis will be addressed in an upcoming work.



XRBs are known to host compact, steady radio jets, typically associated with the canonical low/hard and intermediate/hard states, and are quenched at the high/soft state, where the emission becomes dominated by the standard thin disk \citep{2004fender}. In the AGN context, low luminosity AGNs such as LINERs (Low Ionization Nuclear Emission-Line Regions) often display hard, power-law dominated X-ray spectra and enhanced radio emission relative to the optical-UV. They also lack a prominent big blue bump, in contrast to the thermal disk-dominated spectra typically observed in luminous Seyfert type-1 galaxies \citep{Holius2008ARA&A..46..475H}. Using the diagnostic mid-infrared emission lines of nearby galaxies,   \citet{Ontiveros2021MNRAS.504.5726F} showed that radio-loud LINERs and low-luminosity Seyferts resemble an accretion regime analogous to the \lq hard\rq\, state in XRBs, while radio-quiet Seyfert 1s resemble the \lq soft\rq\, state, providing direct observational support for a common jet-accretion coupling across the black hole mass scale. However, optically bright Seyfert galaxies have been shown to host compact, low-power, or unresolved jets \citep{2019panesa}, reinforcing the relevance of the AGN/XRB analogy beyond classical radio-loud systems. 
Motivated by this framework, we compile archival, quasi-simultaneous radio and X-ray observations of Mkn~590 to investigate the evolution of its radio-to-X-ray luminosity ratio and explore how the radio emission responds to changes in the accretion state.

The structure of the paper is as follows: Sec.~\ref{sec:lab} presents an overview of the source while a rich observational history is presented in the Appendix~\ref{appendixA.1}. The data used in this paper are 
described in Sec.~\ref{sec:data}, ~\ref{sec:radio} and Appendix~\ref{app:1018}. Parametrization of spectral state transition is established in Sec.~\ref{sec:param}, while the Sec.~\ref{sec:alfy}--\ref{sec:fund} present our main results on $\alpha_{\rm ox}$, the hardness-intensity diagrams, and long term radio/X-ray variability respectively. The summary and conclusions are provided in Sec.~\ref{sec:conc} and Sec.~\ref{sec:summary} respectively. 


\section{ Mkn 590: Observations and Data Analysis}
\label{sec:lab}

\begin{figure*}[h]
    \centering
    \includegraphics[width=1.0\linewidth]{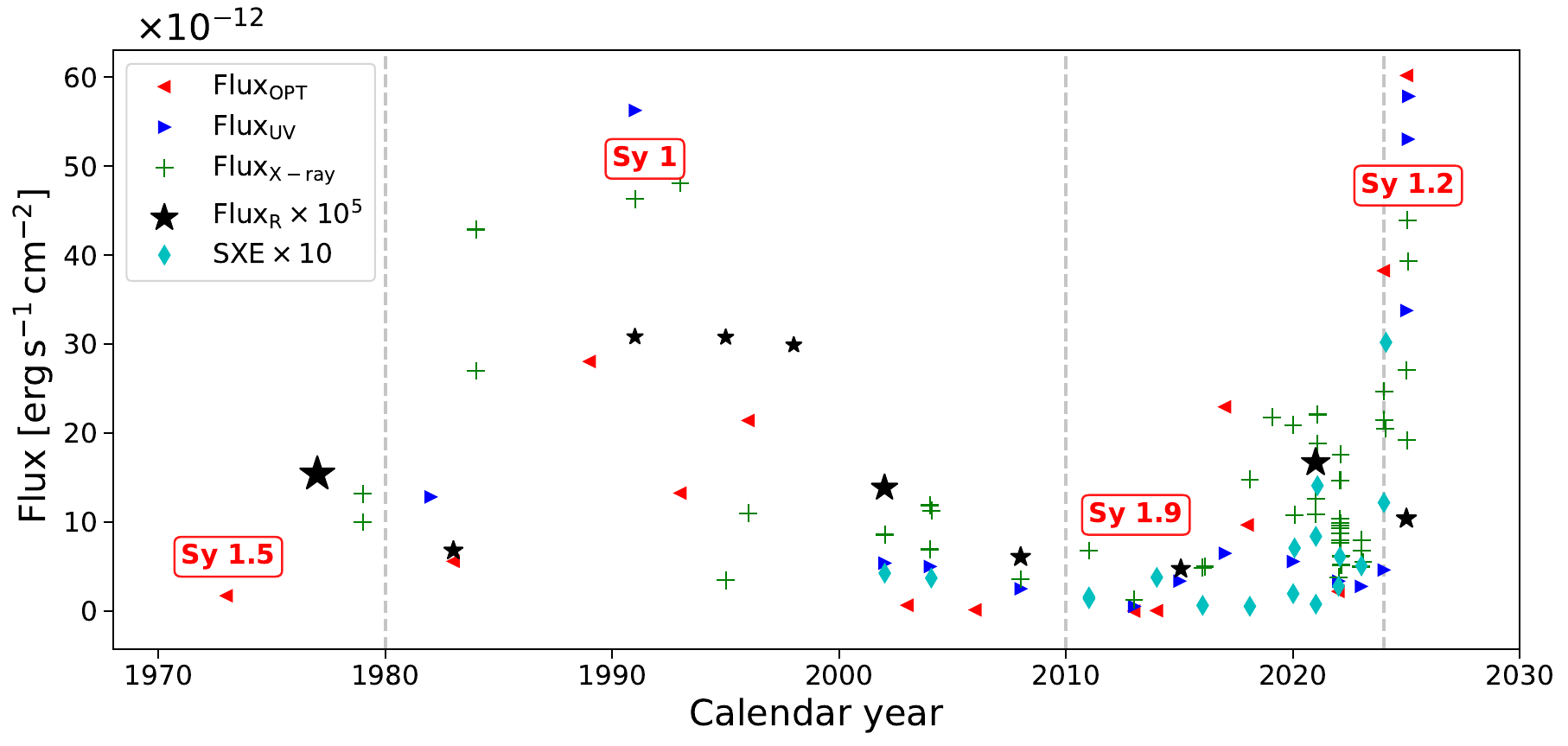}
    \caption{Guideline multi-wavelength light curve of Mkn~590 compiled from various facilities between 1970 and 2025, also presented by \citet{2014Denney,2016koay, lawther2023MNRAS,2025Lawther} and further detailed in Appendix~\ref{appendixA.1}. Details on data collection are described in Appendix~\ref{app:1018} and listed in Tab.~\ref{tab:mrk590_continuum}. 
    In the case of optical, UV and radio measurements we present monochromatic fluxes, while in case of X-rays, the model dependent fluxes integrated over 2--10\,keV are shown. 
    The SXE and radio fluxes are offset vertically for clear comparison with other bands. The marker size of radio data scales with the beam size of radio facility. 
    Three major changing look events in the $\sim$ 1980s, $\sim$ 2010, and $\sim$ late 2023 are  depicted with grey dashed lines.}
    \label{fig:history}
\end{figure*}

Mkn~590 ($z=0.0264$) is one of the best-studied nearby CLAGN, having undergone multiple dramatic spectral and luminosity variations over the past five decades. The source was in a luminous type~1 phase during the late 1980s--1990s, declined into a deep low state by 2012--2015, and has recently entered a new re-brightening episode accompanied by the return of the broad optical emission lines and a sudden re-strengthening of its SXE \citep{2014Denney,Raimundo2017,lawther2023MNRAS,2025Lawther,2025Palit}. 
An overview of its long-term activity is shown in Fig.~\ref{fig:history}, while full details of the historical multi-wavelength activity and data collection are presented in Appendix~\ref{appendixA}. At present, Mkn\,590 remains the subject of extensive multi-wavelength monitoring with a broad range of facilities, alongside ongoing theoretical modeling of its spectral and timing properties. These coordinated efforts aim to probe the structure and evolution of the inner accretion flow as it enters a new elevated accretion regime. 

All data and reduction procedures are described in the following subsections.
For the purpose of our studies we adopt standard cosmological parameters: Hubble constant $H_{\rm 0}$ = 69.6 km s$^{-1}$ Mpc$^{-1}$, mass density $\Omega_{\rm m}$ =0.3, and density of dark energy $\Omega_{\lambda}$=0.7. In estimating the basic observable properties of Mkn\,590, we use a luminosity distance of $L_{\rm D}=115$\,Mpc \citep{NEDW2006PASP..118.1711W}, and the mass of the central black hole $M_{\rm BH}=4.75\, (\pm 0.74) \times 10^7$\,M$_{\odot}$, which has been determined based on extensive reverberation mapping campaigns \citep{2004peterson}. 

\subsection{The {\it Swift} data}
\label{sec:data}

In the last decade, the \textit{Swift} satellite has achieved a total of 420 pointings of Mkn 590, with especially higher cadence observations since 2017, when the source began to exhibit flaring activity (for example, see \citealt{lawther2023MNRAS}).  For the purpose of this study, we analyze the data obtained between December 10, 2013, and September 28, 2025, observed by two instruments onboard \textit{Swift}, the X-Ray telescope (XRT; \citealt{Burrows2005}) and UV/optical Telescope (UVOT; \citealt{roming2005}). This dataset includes bi-weekly cadence observations previously published in the work by \citet{lawther2023MNRAS}, conducted under \textit{Swift} GO Cycle 14 (Programs 1417159 and 1417168; PI: Vestergaard), and joint \textit{NuSTAR-Swift} observations from \textit{NuSTAR} Cycle 5 (Program 5252; PI: Vestergaard). It also incorporates a period of high-cadence monitoring (every 1--2 days) from September 2017 to February 2018, obtained through \textit{Swift} Cycle 13 Director’s Discretionary Time (PI: Vestergaard). The corresponding \textit{Swift} target IDs include 37590, 80903, 88014, 94095, 10949, 11481, 11542, and 13172. We further include data presented by \citet{2025Lawther}, covering the period from June 2021 to September 2024. These data were obtained through \textit{Swift} Cycle 18 (Program 1821134; PI: Lawther), a VLBA-\textit{Swift} joint proposal (Program VLBA/22A-217; PI: Vestergaard), Swift Cycle 19 (Program 1922187; PI: Lawther), and another VLBA-\textit{Swift} joint program (Program VLBA/24A-374; PI: Vestergaard). Additionally, we complemented our entire dataset with new \textit{Swift} observations spanning from between October 2024 to September 2025 (obsIDs 97768014 to 98338027; Program 2124212 cycle 21, PI: Lawther; 
Program 2124218 cycle 21, PI: Walsh;
Program VLBA/25A-323, VLBA/Swift, PI: Walsh).

We retrieved the XRT data products from the UK \textit{Swift} Science Data Center \citep{Evans2009MNRAS.397.1177E}. Each individual XRT spectrum was generated choosing a 30'' circular extraction region centered on the source and a co-spatial background annular region of 35--75''. The UVOT data across all six filters (V, B, U, UVW1, UVM2 and UVW2) were processed using HEASOFT v6.32 and the most recent calibration files. We performed aperture photometry using  \texttt{uvotsource}, employing a 5'' circular region on the source and a 35--75'' co-spatial background annulus. This choice reduces photometric errors which might be arising due to point spread function instability during the orbital motion of the telescope \citep{Poole2008MNRAS.383..627P}. In addition, we neglected observations affected by low sensitivity regions on the detector, resulting in abrupt flux drops \citep{Edelson2019ApJ...870..123E}.

We correct the flux densities in each filter for Galactic extinction using the standard dust maps \citep{extiction1998ApJ...500..525S}. The color excess (E(B-V)= 0.0365) at the source coordinates was obtained assuming an extinction law with $R=3.1$ \citep{cardelli1989ApJ...345..245C}, and we apply extinction corrections at the effective wavelengths of the UVOT filters using the Python packages \texttt{sfdmap}\footnote{\href{https://github.com/kbarbary/sfdmap}{https://github.com/kbarbary/sfdmap}} and \texttt{extinction} \texttt{sfdmap}\footnote{\href{https://extinction.readthedocs.io/en/latest/index.html}{https://extinction.readthedocs.io/en/latest/index.html}}. Finally, we subtract the host galaxy contribution for each filter adopting the methodology followed by \cite{2025Lawther}.

Overall, we obtain 399 XRT, 365 UVOT V, B, U, UVW1, UVW2 and 151 UVM2 data points. The majority of these observations, up to early 2025, have been previously reported \citep{2025Lawther,2025Palit,2025palit_borkar}. In our current work, we incorporate new data obtained through September 28, 2025, during which the average X-ray flux during the month of September 2025 remained elevated at $\sim 7.7 \times 10^{-11}$\,erg\,s$^{-1}$\,cm$^{-2}$ in the 0.3--10 keV band, as shown in Fig.~\ref{fig:history}.

\subsection{Radio and X-ray observations}
\label{sec:radio}

Based on the historical data in Fig.~\ref{fig:history} and Appendix~\ref{appendixA}, we compiled all the contemporaneous X-ray and radio data of Mkn~590 to explore the accretion-ejection behavior. As shown in Tab.~\ref{tab:FP}, we identified seven epochs during which radio and X-ray observations were taken within one year. We converted the radio flux densities measured at different frequencies to a common reference frequency of 5~GHz. To minimize uncertainties arising from differing beam sizes, we restricted this sub-sample to observations with beam sizes of a few arcseconds only. The frequency conversion was performed assuming a standard radio spectral index and intrinsic 1$\sigma$ scatter of $\alpha = -0.8\, (\pm 0.2)$ \citep{condon1992ARA&A..30..575C,2016koay} typical for this source, such that the flux density scales with frequency as $S_{\nu} \propto \nu^{\alpha}$, yielding
\begin{equation}
S_{\mathrm{5\,GHz}} = S_{\mathrm{1.4\,GHz}}
\left( \frac{\nu_{\mathrm{5\,GHz}}}{\nu_{\mathrm{1.4\,GHz}}} \right)^{\alpha}.
\label{eq:radio}
\end{equation}
We retrieved the X-ray flux measurements and upper limits from missions prior to year 2000 (e.g., {\it ROSAT} and {\it Einstein}) from the XMM-Newton upper-limit server interface \footnote{\href{https://www.cosmos.esa.int/web/xmm-newton/uls-userguide}{https://www.cosmos.esa.int/web/xmm-newton/uls-userguide}}. For detections from missions such as \textit{EXOSAT} and \textit{ROSAT}, which typically observed in the 0.2--2~keV band, we converted to the 2--10~keV band using the  \texttt{WebPIMMS}\footnote{\href{https://heasarc.gsfc.nasa.gov/cgi-bin/Tools/w3pimms/w3pimms.pl}{https://heasarc.gsfc.nasa.gov/cgi-bin/Tools/w3pimms/w3pimms.pl}} tool, assuming a power-law continuum with a X-ray photon index of $\Gamma = 2$. Post year-2000, we used \textit{Swift} observations by fitting the X-ray spectra in the 2--10 keV range and extracting the unabsorbed X-ray fluxes. We converted the radio and X-ray flux values to luminosities assuming standard cosmological parameters, and we present those values in Tab.~\ref{tab:FP}. In order to compare the variability trends of Mkn~590 with another classic CLAGN of comparable mass ($M_{\rm BH} = 7\times 10^7 M_{\odot}$), namely Mkn~1018, we adopted the 5 GHz radio and 2--10 keV luminosities from \citet{lyu2021MNRAS.506.4188L}. They are presented in Tab.~\ref{tab:lyu}.

\begin{table}[h]
\centering
\caption{An overview of the contemporaneous radio and X-ray fluxes of Mkn~590 to study the correlated radio/X-ray behavior (described in Sec.~\ref{sec:radio}). The computed luminosities in the radio and X-rays are shown in columns 3 and 4, respectively, followed by their ratio given in column 5, and representative year in column 6.}
\label{tab:FP}
\begin{tabular}{|c|c|c|c|c|c|}
\hline
 Radio facility & X-ray mission &  \multirow{ 2}{*}{$\log \left( \frac{\nu L_{\rm 5\,GHz}}{{\rm erg}\, {\rm cm}^{-2}} \right)$}  &  \multirow{ 2}{*}{$\log \left( \frac{L_{\rm 2-10\,keV}}{{\rm erg}\, {\rm cm}^{-2}} \right)$}  & 
  \multirow{ 2}{*}{$\log\left(\frac{\nu L_{\rm 5\,GHz}}{ L_{\rm 2\text{-}10\,keV}}\right)$} & Representative \\
  5 GHz flux density  & 2-10 keV Flux &  &   &   &   \\
  (mJy) & ($\times 10^{-11}$ erg\,s$^{-1}$\,cm$^{-2}$) &  &   &    & Year\\
  (1) & (2) & (3) & (4) & (5) & (6)\\
     \hline \hline
     VLA-A & \textit{EXOSAT}   & 
     \multirow{ 2}{*}{$38.14 \pm 0.12$} &     \multirow{ 2}{*}{$ 43.75 \pm 0.01$} &  \multirow{ 2}{*}{$-$5.61}   & \multirow{ 2}{*}{1984}   \\ 
 1.75 $\pm$0.49 &3.63 $\pm$ 0.11  & &  & &  \\
\hline
VLA-A & \textit{ROSAT}  & \multirow{ 2}{*}{$38.40 \pm 0.05$} &    \multirow{ 2}{*}{$43.77 \pm 0.01$} &  \multirow{ 2}{*}{$-$5.37}    & \multirow{ 2}{*}{1991}   \\
 5.55 $\pm$ 0.62 &3.82 $\pm$ 0.09 & & & &  \\
\hline
MERLIN & \textit{ROSAT} &  \multirow{ 2}{*}{$ 38.52 \pm 0.03$}  & \multirow{ 2}{*}{$ 42.74 \pm 0.01$} &   \multirow{ 2}{*}{$-$4.13}      & \multirow{ 2}{*}{1995} \\ 
 4.16 $\pm$ 0.30 &0.286 $\pm$ 0.01  & & & &  \\
\hline
VLA-A & \textit{Swift}-XRT & \multirow{ 2}{*}{$38.09 \pm 0.11$} & \multirow{ 2}{*}{$ 42.65 \pm 0.04$}  & \multirow{ 2}{*}{$-$4.66}    & \multirow{ 2}{*}{2008}     \\ 
 1.57 $\pm$ 0.40 &0.36 $\pm$ 0.03 & & & &  \\
\hline
VLA-A & \textit{Swift}-XRT & \multirow{ 2}{*}{ $37.96 \pm 0.14$} & \multirow{ 2}{*}{ $ 42.58 \pm 0.08$ } & \multirow{ 2}{*}{$-$4.62} &  \multirow{ 2}{*}{2015}   \\ 
 1.15 $\pm$ 0.31& 0.24 $\pm$ 0.05  & & & &  \\
\hline
VAST &  \textit{Swift}-XRT & \multirow{ 2}{*}{ $38.50$} & \multirow{ 2}{*}{$43.24 \pm 0.02 $} & \multirow{ 2}{*}{$-$4.74}  &  \multirow{ 2}{*}{2021}  \\ 
 4.03  &  1.10 $\pm$ 0.05& & & &  \\
 \hline
GMRT &  \textit{Swift}-XRT &  \multirow{ 2}{*}{ $38.32 \pm 0.12$}  & \multirow{ 2}{*}{$ 43.60 \pm 0.05 $} &  \multirow{ 2}{*}{$-$5.28}    &  \multirow{ 2}{*}{2025}    \\ 
 2.68 $\pm$ 0.69& 2.54 $\pm$ 0.30 & & & &  \\
\hline
\end{tabular}
\end{table}

\section{Parametrization of spectral state transitions}
\label{sec:param}
We wish to probe the changes in the spectral energy distribution of Mkn 590 so to examine if an actual state change takes place over the time span of our database. We do this by following the evolution of key AGN diagnostic parameters such as the X-ray loudness parameter $\alpha_{\textsc{ox}}$, the Eddington fraction 
$\eddfrac$, the hardness ratio $HR$, and the radio-to-X-ray scaling. In this section, we describe the methods used to estimate these quantities from the observational data.

The classical $\alpha_{\textsc{ox}}$ for AGN and quasars is given by
\begin{equation}
    \aox=-\frac{\log \left[L_{\rm 2 keV}/L_{\rm 2500 \AA\ } \right] }{2.605}\,,
    \label{eq:alphaox1}
\end{equation}
\citep{1979tananbaum,2009sobolewska}, with
2500\,{\AA } and 2\,keV emission as proxies for emission from the UV-emitting disk and the X-ray corona, respectively. \citet{2025Palit} found that the re-brightening of Mkn~590 is accompanied by the detection of the SXE, potentially present below 2\,keV. Thus, for the purpose of this paper, we define two new $\aox$  parameters,
corresponding to the soft and the hard X-ray bands,
which better reflect the importance of the warm and hot corona, respectively, relative to the cold disk: 
\begin{equation}
    \aoxone=-\frac{\log \left[L_{\rm 0.5 keV}/L_{\rm 2500 \AA} \right ]}{2.003} \,,
     \label{eq:alphaox2}
\end{equation}

\begin{equation}
    \aoxtwo=-\frac{\log \left[ L_{\rm 4 keV}/ L_{\rm 2500 \AA} \right] }{2.906} \,.
     \label{eq:alphaox3}
\end{equation}

For all three versions of the $\aox$, we estimate the monochromatic luminosity at 2500~\AA, after selecting only those \textit{Swift}-UVOT observations with simultaneous UVW1 and UVW2 coverage. After subtracting the host galaxy contribution and correcting the photometric flux densities for Galactic extinction, as described in Sec.~\ref{sec:data},
we use a simple linear interpolation to estimate the flux density at 2500~\AA\ and convert it into monochromatic luminosity.
Next, to derive monochromatic X-ray fluxes,
we fit the \textit{Swift}-XRT spectra using a 
\lq pegged powerlaw model\rq\, (\texttt{pegpwrlw}) in \texttt{XSPEC}  \citep[version 12.13.1; ][]{arnaud1996}. 
The direct way to get a monochromatic flux with this model is to set par2=par3 in $\texttt{XSPEC}$, yielding the flux in mJy at that energy.
For the monochromatic 0.5~keV fluxes, we fit across the 0.3--2~keV band only;
for the 2 and 4~keV monochromatic fluxes, we
fit across the 2--10\,keV band.
We then convert these monochromatic fluxes into monochromatic luminosities 
in units of erg\,s$^{-1}$\,Hz$^{-1}$. 

We parametrize Mkn~590's highly variable accretion rate by the ratio of bolometric luminosity to Eddington luminosity, i.e. $\eddfrac=L_{\rm bol}/L_{\rm Edd}$. There are two main uncertainties associated with estimating $\eddfrac$, namely the black hole mass ($M_{\rm BH}$), and bolometric correction factor ($K_{\rm bol}$), defined as the ratio of $L_{\rm bol}$ to X-ray luminosity 
$L_{\rm X}$. Several distinct formalisms have been developed so far to empirically estimate $K_{\rm bol}$ based on the 
relatively stable accretion phases of AGN and quasars
(for discussion see  Appendix~\ref{appendixC.1}). However, as shown in Fig.~\ref{fig:history} and further demonstrated in this work, Mkn~590 has undergone dramatic accretion rate transitions, exhibiting strong variability in both the X-ray and optical/UV bands. As demonstrated in Appendix~\ref{appendixC.2}, the choice of $L_{\rm bol}$ estimator significantly affects the inferred $\alpha_{\textsc{ox}} -\eddfrac$ relationship. This motivates the use of a source-specific approach: namely, directly integrating the intrinsic broadband SED over 0.001--50 keV (further detailed in Appendix~\ref{appendixC.1}). In this work, we adopt the SED integration method for the primary scientific analysis since it provides the most accurate measure of the 
bolometric luminosity of Mkn~590's nuclear emission.

Another widely used diagnostic for spectral states in X-ray studies of XRBs is the X-ray hardness ratio ($HR$). It provides a  model-independent measure of the X-ray spectral shape. We adopt the definition
$HR=(H-S)/(H+S)$, where $H$ and $S$ are the 2--10~keV and 0.3--2~keV photon count rates, respectively.
In XRBs, they represent the X-ray corona and the thermal disk emission respectively, hence, $HR$ acts as a proxy for the corona-dominated versus the disk-dominated spectral states. 
In AGN, however, the accretion disk emits predominantly in the EUV rather than in soft X-rays, so in this case
$HR$ traces mainly changes within the hot/warm coronal components. We use 
$\aoxone$ and $\aoxtwo$ to track the ratio of the power between the disk and warm or hot coronae, respectively.

We propagated the 1$\sigma$ uncertainties from UVOT photometry, X-ray spectral fitting, observed XRT count rates, and the reported errors in archival radio and X-ray measurements into the final uncertainties on the derived quantities $\alpha_{\textsc{ox}}$, $HR$, and the radio-to-X-ray luminosity ratio. This was done by adding the relative errors of the individual measurements in quadrature (i.e., standard error propagation). We followed the same strategy to estimate the uncertainties on $\eddfrac$.

\section{The observed $\aox$--$\eddfrac$ relation}
\label{sec:alfy}

\begin{figure*}[h]
    \centering
    \includegraphics[width=.99\linewidth]{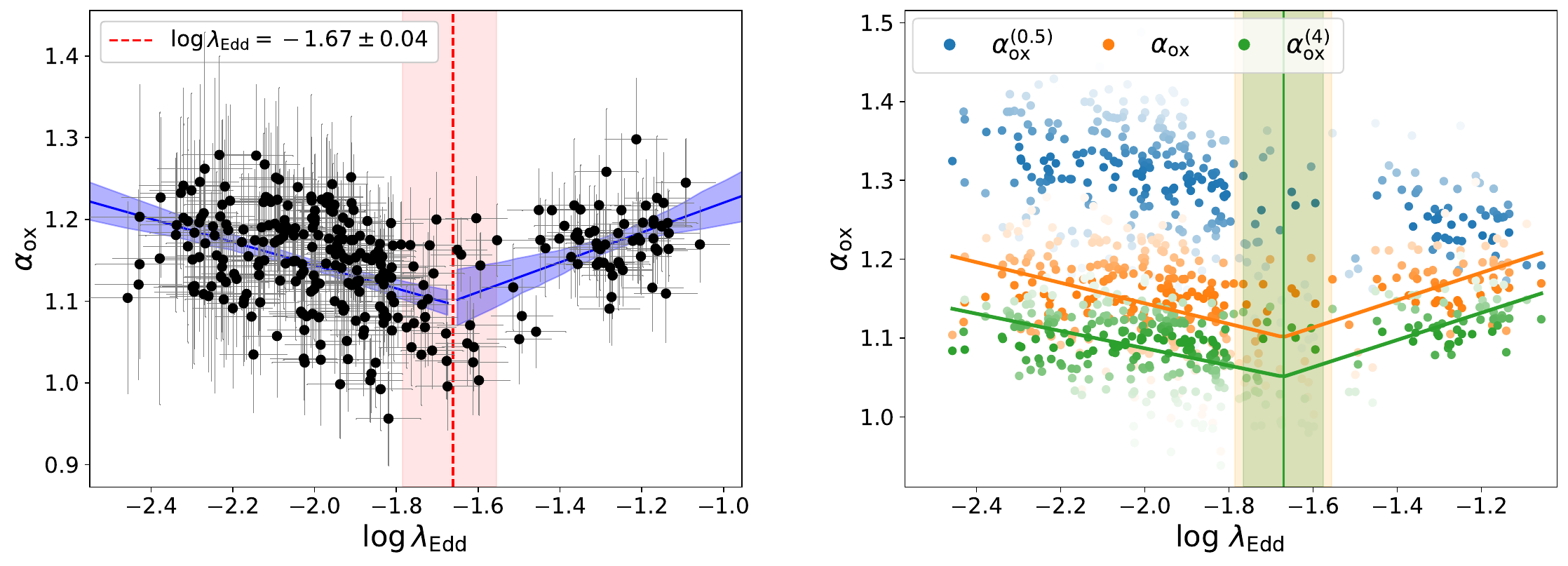}
    \caption{{\it Left}: the variation of $\aox$ with $\eddfrac$ from {\it Swift} monitoring of Mkn~590 from Dec.\ 10, 2013 to Sept.\ 28, 2025. The scatter of data exhibits a \lq V\rq-shaped trend, with a break point indicated by the vertical red dashed line. The associated 95\% confidence interval is shown as a red shaded band. Linear fits on either side of the break point are indicated by solid blue lines, while their 95\% confidence region are shown in purple shading. The correlation strengths, estimated using the Spearman rank coefficient, are $\rho = -0.38$ for the left branch and $\rho = 0.37$ for the right branch, with both relations rejecting the null hypothesis at high significance ($p << 10^{-10}$). {\it Right} panels: the same variation, but for $\aoxone$ shown in blue, $\aox$ in orange, and $\aoxtwo$ in green points. Points are color coded by point density to highlight regions of higher concentration and improve visualization of the underlying trend. In the case of $\aoxtwo$, a turnover was detected at $\log \eddfrac= -1.67 \pm 0.04$, a value identical to that for the $\aox$ relation and is shown by vertical lines together with their confidence intervals as shaded regions. The values of $\rho$ for left and right branches of $\aoxtwo$ are  $-0.42$ and $0.45$ respectively, $p << 10^{-10}$. In the case of $\aoxone$, no such break point was found.}
    \label{fig:alpha_ox1}
\end{figure*}

The evolution of $\aox$ as a function of $\eddfrac$ in Mkn~590 marks the most detailed measurement of $\aox$ in a single CLAGN across an entire changing look event.
This is shown in the left panel of Fig~\ref{fig:alpha_ox1}, tracing a clear 
\lq V\rq-shape.
In order to determine the turnover point between the two arms in the \lq V\rq-shaped trend, we employed a piecewise linear regression in Python\footnote{\href{https://github.com/cjekel/piecewise\_linear\_fit\_py}{https://github.com/cjekel/piecewise\_linear\_fit\_py}} \citep{Pilgrim2021}. 
The algorithm (described in Appendix~\ref{appendixD}) detects only one statistically significant break point at $\log \eddfrac = -1.67 \pm 0.04$ with a 
null hypothesis probability of $p << 10^{-10}$. 
The position of the break point is marked by a vertical, red, dashed line in Fig.~\ref{fig:alpha_ox1}, and the 95\% confidence interval is indicated by a light red shaded region. 
Subsequently, we fitted the data on each side of this break point using Deming regression, a statistical technique that accounts for measurement uncertainties in both variables and estimates the best fit line by minimizing the orthogonal (error-weighted) distances of the points from the regression model (outlined in Appendix~\ref{appendixD}). To estimate the confidence intervals for Deming regression, we performed 5000 bootstrap resamplings, deriving the 95\% confidence regions shown as purple shaded bands in Fig.~\ref{fig:alpha_ox1}. The best fit linear models for the left and the right branches are $\aox = (-0.14 \pm 0.02)\log \eddfrac+(0.86 \pm 0.04 )$ and $\aox = (0.18 \pm 0.05)\log \eddfrac+(1.40 \pm 0.06) $, respectively. 

\begin{figure}[h]
\centering
    \includegraphics[scale=.40]{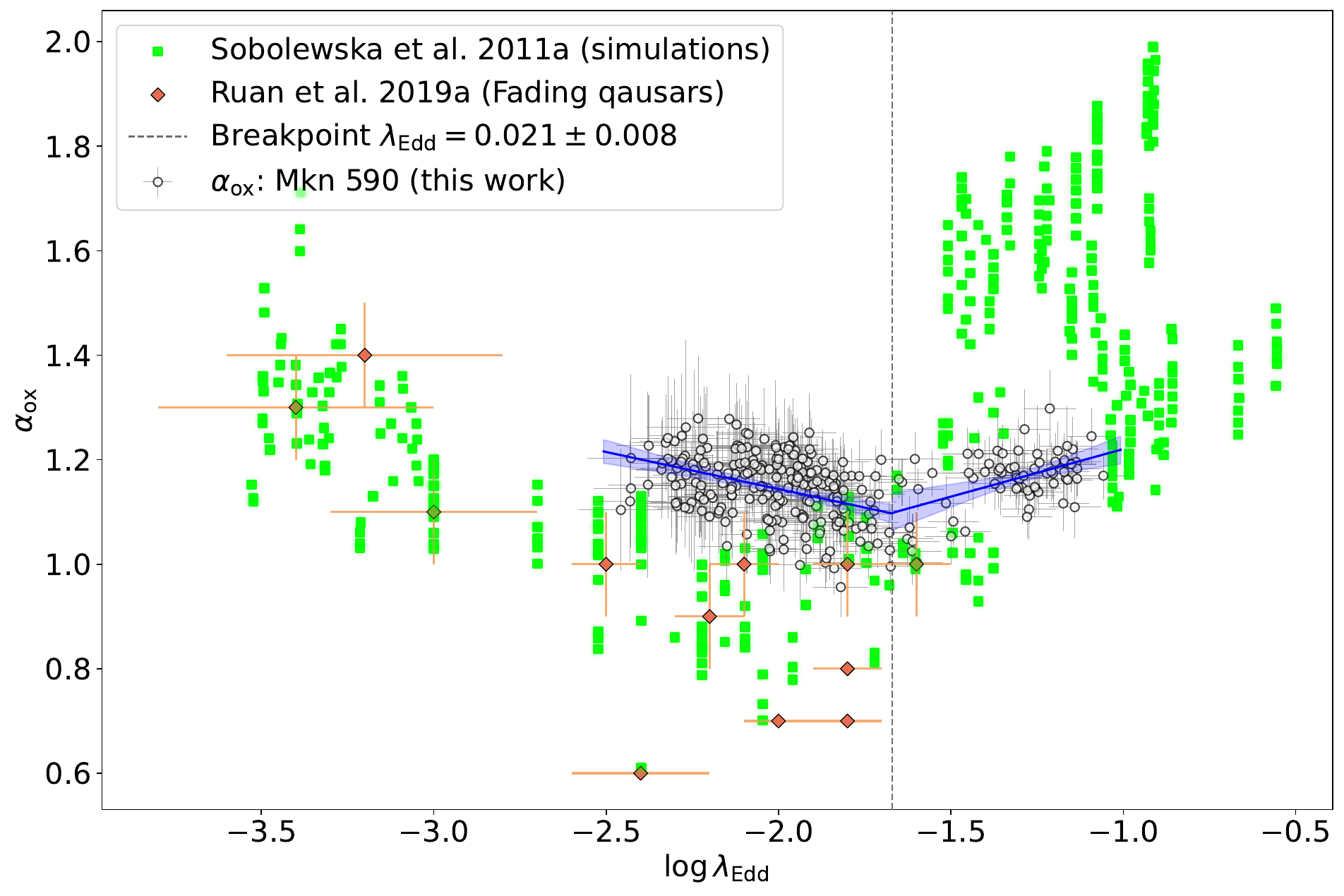}
    \caption{A comparison of the observed $\aox$--$\eddfrac$ relation (with $L_{\rm bol}$ computed by the SED integration method) for Mkn~590 given by open circles, with observations of several fading quasars marked by red diamonds \citep{Ruan2019ApJ...883...76R}, and with 
    simulated predictions for AGN marked by green squares \cite{Sobo2011MNRAS.413.2259S}. The vertical dashed line represents the break point of the observed relation (see Fig.~\ref{fig:alpha_ox1}), while blue lines show linear fits to the data, with shaded regions indicating 1$\sigma$ confidence intervals.}
    \label{fig:grand_aox}
\end{figure}

At about 2\% of $L_{\rm bol}$, the break point is comparable to the empirical critical Eddington ratio around which changing look events have been observed in AGNs, quasars, and tidal disruption events \citep{Noda2018MNRAS.480.3898N, 2025hagen,jana2025A&A...693A..35J,2025gilbert,duffy2025ApJ...989..102D,wevers2021ApJ...912..151W}. In \citet{Sobo2011MNRAS.413.2259S}, the authors simulated values of $\alpha_{\rm ox}$ for SMBHs in the black hole mass range $10^{6-8}~M_{\odot}$ by scaling the observed spectral state evolution of the XRB GRO J1655$-$40. They recovered the characteristic ‘V’-shaped dependence of $\alpha_{\rm ox}$ 
on Eddington ratio. In Fig.~\ref{fig:grand_aox}, we compare our result for  Mkn~590 with their simulated tracks of $\aox$ (shown as green squares) and  with a sample of quasars that have undergone changing-look events from  bright to faint flux states (hence fading quasars;  \citealt{Ruan2019ApJ...883...76R}; shown as red diamonds). The simulated $\aox$ parameter exhibits substantially larger scatter than the Mkn~590 data, as they span an ensemble of black hole masses. Furthermore, the larger vertical scatter on the high-$\eddfrac$ branch likely arises from the coexistence of both canonical hard and soft accretion states above the critical Eddington ratio in the simulations. The comparison with our results is therefore intended only to test the qualitative behaviour, in particular the turnover in the $\aox - \eddfrac$ relation at $\eddfrac \sim 0.01$. As indicated, fading quasars predominantly occupy the left branch of this trend, consistent with the simulated low/hard state of AGNs while Mkn~590 spans both sides of the break. As we show in the next section, the pre-changing look phase populates on the left branch and the post-changing look phase populates the right branch.

Next, we compare
the dependence of $\aoxone$, $\aox$ and $\aoxtwo$ on $\eddfrac$ in the right panel of Fig.~\ref{fig:alpha_ox1}.
The widespread presence of the warm corona covering the accretion disk in Mkn~590 \citep{2025Lawther} can significantly modulate the overall shape of the SED, which is expected to be probed by the newly defined $\aoxone$. The absence of a clear \lq V\rq-shaped trend and the lack of detected break points in the $\aoxone$--$\eddfrac$ relation clearly highlight its 
diverging 
behavior compared to the standard $\aox$. The results suggests a tight radiative and possibly structural coupling between the warm corona and the UV emitting disk as discussed later.  On the other hand, the $\aoxtwo$ parameter exhibits the expected \lq V\rq-shaped trend similar to that displayed in the left panel. This indicates that the 2 and 4 keV fluxes are dominated by the hot coronal power-law component and are not significantly contaminated by the SXE emission. 


\begin{figure}[h]
    \centering
    \includegraphics[width=.99\linewidth]{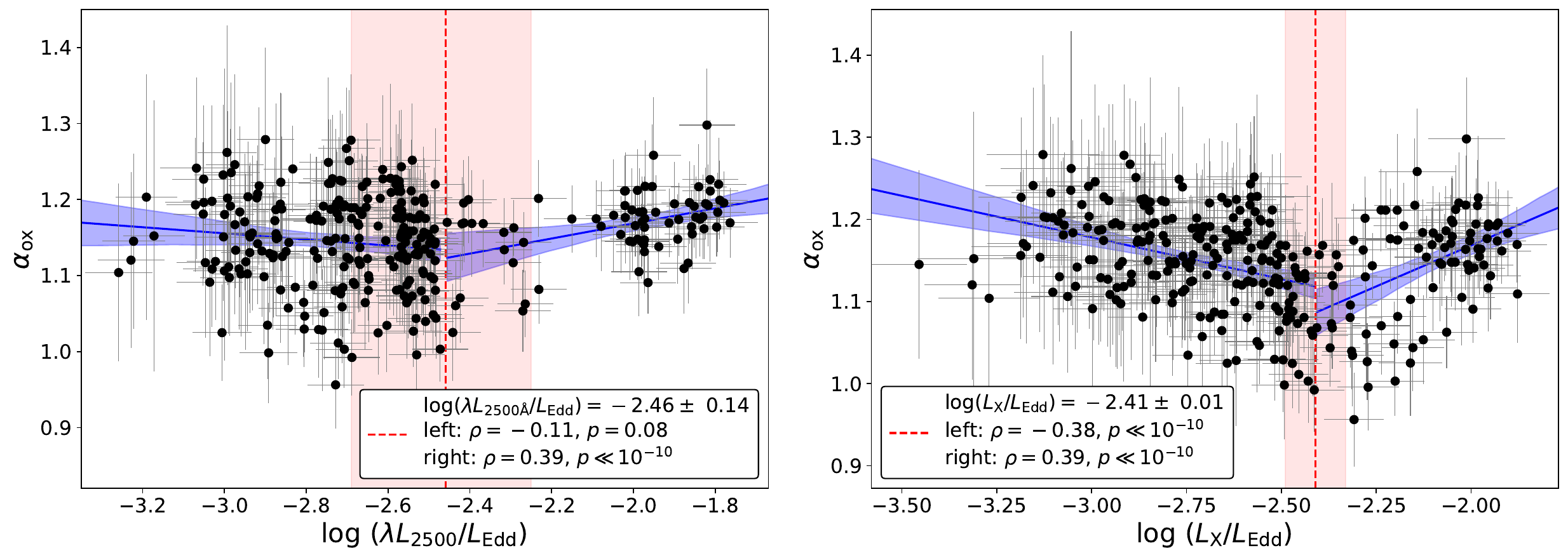}
    \caption{The dependence of $\aox$ on the UV and hard X-ray accretion power tracers
    $\lambda L_{2500}/L_{\rm Edd}$ and
    $L_{\rm X}/L_{\rm Edd}$. The Spearman correlation strength ($\rho$) and null hypothesis (p-values) are reported for each arm around the break-point (red dashed line).
     The meaning of colors is the same as in the left panel of Fig.~\ref{fig:alpha_ox1}.}
    \label{fig:lumfrac}
\end{figure}

In order to test the robustness of the observed $\aox$ evolution against the choice of tracers of accretion power, we use the ratio of $\lambda L_{2500}/L_{\rm Edd}$, which primarily traces thermal emission from an accretion disk in UV, and  the ratio of $L_{\rm X}/L_{\rm Edd}$, where $L_{\rm X}$ is the integrated 2--10\,keV luminosity representing the power from the hot corona. It must be noted that we do not impose any link between the $L_X$ and $L_{\rm 2500}$ luminosities a priori. The dependence of $\aox$ on the UV and X-ray tracers is shown in the left and right panels of Fig.~\ref{fig:lumfrac}, respectively. 
Again, a \lq V\rq-shaped relationship persist for both tracers, with single break-points detected at high significance (p-value $< 10^{-10}$). The best fit linear models for the left and right arms of the $\aox-\log(\lambda L_{2500}/L_{\rm Edd})$ relation are $\aox = (-0.05 \pm 0.03) \log(\lambda L_{2500}/L_{\rm Edd}) + (0.98 \pm 0.07) $ and $\aox = (0.09 \pm 0.02) \log(\lambda L_{2500}/L_{\rm Edd}) + (1.34 \pm 0.04) $ respectively. And, for the $\aox-\log(L_{X}/L_{\rm Edd})$  relation, $\aox = (-0.15 \pm 0.02)\log(L_{X}/L_{\rm Edd}) + (0.73 \pm 0.05) $ and $\aox = (0.25 \pm 0.04)\log(L_{X}/L_{\rm Edd}) + (1.67 \pm 0.08) $ respectively. It is worth noting that the linear relationship based on the UV accretion power tracer is shallower (and consistent with zero) than the reference relation of $\aox - \eddfrac$ in Fig.~\ref{fig:alpha_ox1}, while the relations based on X-ray accretion have comparable slopes to that of the reference one. Nevertheless, both the relationships in Fig.~\ref{fig:lumfrac} exhibit comparable values for the turnover, with statistically significant break points at $\log(\lambda L_{2500}/L_{\rm Edd}) = -2.46 \pm 0.14$ and $\log(L_{\rm X}/L_{\rm Edd}) = -2.41 \pm 0.01$, respectively, as indicated by the vertical dashed lines and shaded regions. As an additional consistency check, we examined the direct relation between the UV- and X-ray-based Eddington-ratio tracers (further detailed at the end of Appendix~\ref{appendixC.2}). These quantities are themselves strongly correlated throughout the monitoring campaign (Spearman $\rho$=0.90, p-value $\ll 10^{-10}$). The piecewise linear regression revealed a statistically significant break at $\log(\lambda L_{2500}/L_{\rm Edd}) = -2.44 \pm 0.10$. This is consistent with the two break locations inferred above.
\section{Hardness--Intensity Diagrams}
\label{sec:hids}

\begin{figure}[h]
    \centering
    \includegraphics[width=0.97\linewidth]{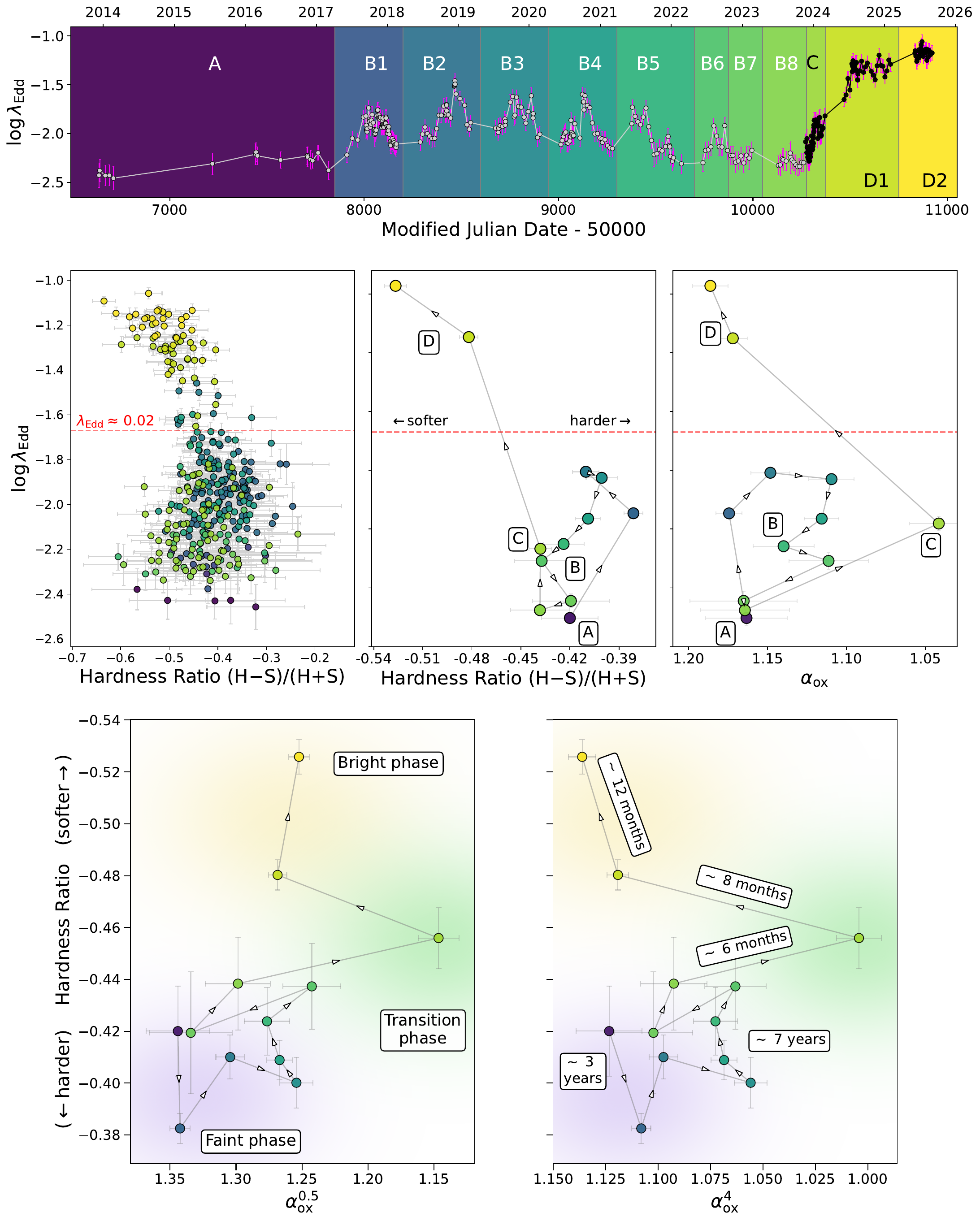}
    \caption{{\it Top}: Temporal variation of $\eddfrac$, segmented into slots named from A to D2, and color coded for reference to the following sub-panels below. {\it Middle}: From left to right: $\eddfrac$ versus $HR$ for all \textit{Swift} data, the same but binned across color coded time slots, and  $\eddfrac$ versus $\aox$ relation binned in the same way, respectively. 
    \textit{Bottom:} Binned $HR$ vs $\aoxone$ (left) and $\aoxtwo$ (right) relations both show the coupled behavior of the UV thermal disk with the warm corona, and the coupling between the warm and hot coronal components, respectively. The three broad phenomenological regions occupied by the source are shaded to guide the eye. The approximate timescales correspond to epochs A $\rightarrow$ B1, B1$\rightarrow$ B8, B8 $\rightarrow$ C, C$\rightarrow$D1 and D1$\rightarrow$D2, respectively. }
    \label{fig:HID}
\end{figure}

In an attempt to identify distinct accretion regimes in this CLAGN, we examine the hardness-intensity diagrams (hereafter HIDs) of Mkn~590. This exercise is not meant as a direct comparison
to XRBs, since the observables used here are not direct analogues of canonical XRB state diagnostics, but is meant as phenomenological only. Moreover, the source does not exhibit a large dynamic range in luminosity and spectral hardness that is typically seen during full XRB state cycle. Moreover, we lack observations of Mkn~590 during $\sim 2008-2012$ which correspond to the faintest X-ray fluxes \citep{2012rivers}.
The top panel of Fig.~\ref{fig:HID} presents the temporal variations in $\eddfrac$, derived from the \textit{Swift} data (defined in Sec~\ref{sec:param}). This sequence is segmented into broader phases (A--D) and finer sub-epochs (e.g., B1--B8), each color coded for consistency across subsequent analyses. The sharp increase in $\eddfrac$ by $ \Delta \eddfrac \sim 1.2$ dex starting late in year 2023 
marks the beginning of the changing look event \citep{2025Palit} and seems to trace the turnover point in the \lq V\rq\,-shaped trends shown previously.  

A more detailed examination of the interplay between the emission components is presented in the middle row of Fig.~\ref{fig:HID}, which shows the evolution of $HR$  (left and center panels), and $\aox$ (right panel) as a function of $\eddfrac$. The data points are color coded according to the temporal segments defined in the top panel of the figure. The red dashed line indicates the break point in the $\aox$-$\eddfrac$ relation determined in Sec.~\ref{sec:alfy}. In the leftmost $\eddfrac$-$HR$ plot, the source traces a broadly concave trajectory upwards, reminiscent of the characteristic color-color diagrams observed in XRBs \citep[][and references therein]{kalemci2022hxga.book....9K}, where accreting systems generally evolve from canonical hard-to-soft states with increasing luminosity. For better visualization of temporal evolution, the data in the middle and right panels are binned by segments, with arrows indicating the direction of progression through time. Starting from the faintest state (A), it traced out an irregular loop (B1--B8) while remaining restricted to relatively harder values of $HR$, and with variability temporarily following a harder-when-brighter trend. For the same epochs, $\aox$ also traced out an irregular loop (B1--B8) evolving towards harder values of $\aox$, indicating a relative dominance of hard X-rays over the thermal UV emission. Historically, as well as shown in the top panel, this period (B) is marked by the characteristic flaring activity seen in both optical, UV and X-rays, before gradually approaching the pre-flare dip at B8. 

A major contrast appears in these plots for the point of transition between regions B8 and C, where at consistent value of $HR \approx 0.45$ the $\aox$ drops from $\sim 1.17$ to $\lesssim 1.05$ (change by $\gtrsim 0.32$ dex), corresponding to previously presented turnover point. Between B8--C, the $\eddfrac$ varies only by $\lesssim 0.2$ dex, yet the source exhibits a pronounced suppression of the thermal UV emission relative to the X-ray component. From epoch \lq C\rq\ onward, the source rapidly evolves along a distinct branch, characterized by the softer values of both $HR$ and $\aox$, thus indicating a potential switch to a softer-when-brighter behaviour, and also populating the right side of the \lq V\rq\,-shape in Fig.~\ref{fig:alpha_ox1}. We argue in Sec.~\ref{sec:HID_conc} that this switch signals a change of disk-to-corona structure.

In the bottom row of Fig.~\ref{fig:HID}, we further explore the coupling between UV disk and warm or hot X-ray regions by plotting $HR$ against the two new X-ray loudness parameters $\aox^{(0.5)}$ and $\aox^{(4)}$. We find that Mkn~590 occupies three regions in this parameter space, highlighted by shaded areas to guide the eye. The meaning of those trajectory plots is that towards the lower left 
regions of the plots (purple region), we expect relatively harder X-ray emission,  with higher $HR$ values denoting dominance of hot over warm corona, together with relatively stronger UV disc emission i.e. higher value of each $\aox$ parameter. 
Epochs 2014--2017 (A; Faint phase), corresponding to the lower-flux state in both UV and X-rays, primarily occupy this region. A gradual movement towards the right means that UV disc emission gets relatively weaker in comparison to the X-rays  (lower value of each $\aox$ parameter). The flaring epochs between 2017--2023 (B) largely trace this path as the source increased its overall luminosity. The rightmost region (green) marks the transitional phase C, occurring at intermediate $HR$ but characterized by comparatively weak UV emission (lower $\aox$). Finally, the movement towards the upper left of the plot (yellow) indicates both a softening of the overall X-ray spectrum and a recovery of the UV emission. This corresponds to epochs 2024-2025 (D) when the $\eddfrac$ increased $\sim 1.5$ orders of magnitude from the faint phase. In both plots, the $HR$ softens considerably during the bright phase, indicating the domination of soft over hard X-rays, while $\aoxtwo$ reaches the same levels as in its faint phase unlike  $\aoxone$. This is a strong indicator of the warm corona becoming energetically important, and likely coupled to the strong UV disk emission in this phase. Finally, from the bottom rightmost panel, one can see that the source  passes through multiple accretion activity phases over timescales ranging from months to years, driven probably by a complex interplay between the disk and warm  and hot coronae. We further discuss these in the context of the source evolution and its state transition in Sec.~\ref{sec:conc}.

\section{{\bf The radio/X-ray correlation}}
\label{sec:fund}
\begin{figure}[h]
    \centering
    \includegraphics[width=.95\linewidth]{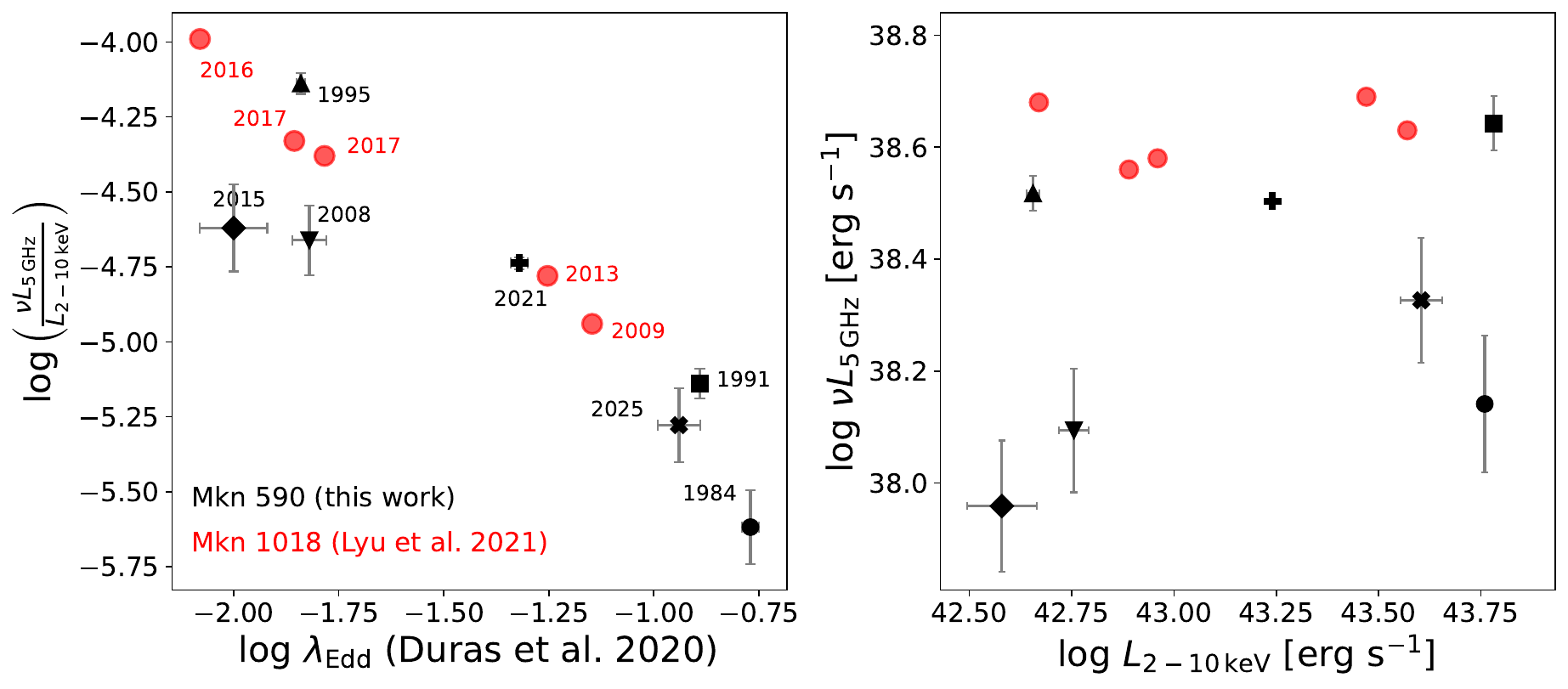}
    \caption{\textit{Left}: The variation of the radio-to-X-ray luminosity ratio measured for Mkn~590 (black symbols; Tab.~\ref{tab:FP}) and for Mkn~1018 (red solid circles; Tab.~\ref{tab:lyu} and Appendix~\ref{app:1018}). \textit{Right}: Radio versus X-ray luminosity for the same epochs as shown in left panel. Identical marker symbols are used in both panels to denote matching observational epochs.}
    \label{fig:radio}
\end{figure}
We examine the changes of the radio and X-ray emission since year 1984. The radio and X-ray  luminosities of Mkn~590 are listed in Table~\ref{tab:FP}. To estimate the Eddington fraction, we adopt the hard X-ray bolometric correction of \citet{Duras2020A&A...636A..73D}, which enables a uniform determination of $L_{\rm bol}$ across epochs lacking full SED coverage (discussed in Appendix~\ref{appendixC.1}). The left panel of Fig.~\ref{fig:radio} shows the ratio of radio to X-ray luminosity as a function of $\eddfrac$ for Mkn~590 (black symbols), with representative epochs annotated. The data suggest a decline in radio power relative to the X-rays with increasing $\eddfrac$. We note that a similar decline in the radio to X-ray luminosity is seen for another CLAGN, Mkn 1018 (red circles in Fig.~\ref{fig:radio}). The data for Mkn~1018 are taken from \citet{lyu2021MNRAS.506.4188L}, while the procedure to compute the corresponding quantities is outlined in Tab.~\ref{tab:lyu} and Appendix~\ref{app:1018}.

 The right panel of Fig.~\ref{fig:radio} compares the radio and X-ray luminosities of Mkn~590, together with those of Mkn~1018 \citep{lyu2021MNRAS.506.4188L}. In both sources, we do not find a statistically significant correlation between $L_{\rm 5GHz}$ and $L_{\rm 2-10keV}$. Although intrinsic cm-band variability has previously been reported for both objects \citep{2016koay,lyu2021MNRAS.506.4188L}, the radio luminosity varies over a much smaller dynamic range than the X-ray luminosity. We caution, however, that the radio measurements presented here for Mkn~590 were obtained with different instruments and different beam sizes of the order of few arcseconds which may introduce additional scatter in the measured radio luminosities beyond the intrinsic source variability.

\section{Discussions}
\label{sec:conc}
For decades, the effort to unify accretion physics across the wide black hole mass range from stellar mass black holes in XRBs to SMBH in AGNs has represented one of the central challenges in high energy astrophysics. However, the discovery of extreme AGN variability, particularly in CLAGN, made a breakthrough this pursuit. Now these rare sources offer a unique window into the dynamic evolution of accretion flow onto SMBH, on observationally accessible timescales. In this study, we investigate a changing look event in the AGN Mkn~590, searching for evidence of accretion state transitions analogous to those observed in XRBs. Utilizing comprehensive multi-epoch monitoring from the  \textit{Neil Gehrels Swift} Observatory, along with nearly fifty years of archival multi-band data, we conduct a series of diagnostic analyses to evaluate the temporal evolution of Mkn~590 in the context of its spectral state transition. 

\subsection{Evidence for an XRB-like state transition}
\label{sec:aox_conc}

Mkn~590 shows a clear \lq V\rq-shaped trend with a statistically robust break point at $\eddfrac = 0.021 \pm 0.008$. This value is broadly consistent with break points reported in other accreting compact objects (Fig.~\ref{fig:grand_aox}). Such behavior has been interpreted as evidence for a transition in the inner accretion flow, from a geometrically thick, optically thin advection-dominated accretion flow (ADAF; \citealt{Narayan1995ApJ...452..710N,Esin1997ApJ...489..865E}) below $\eddfrac \lesssim 0.01$, to a standard thin disk at higher accretion rates ($ 0.01 \lesssim \eddfrac \lesssim 0.1$). 

The origin of the \lq V\rq-shaped trend can be understood based on a truncated accretion disk (\citet{2007Done} present a general overview). From moderately low toward higher Eddington ratios, the left branch of the trend reflects a regime in which the optically thick disk is likely truncated or recessed, providing a limited supply of UV seed photons to the hot corona. In this configuration, the Comptonized X-ray emission dominates over the thermal disk emission, leading to a harder spectrum and a lower $\aox$ value. Such behavior is consistent with trends observed in XRBs during their canonical hard and hard-intermediate states, where spectral hardening accompanies increasing luminosity \citep{sobo2011MNRAS.417..280S, connoly2016MNRAS.459.3963C}. At even lower Eddington ratios ($\eddfrac \ll 0.003$), although not fully sampled in the data available to date, the observed increase in $\aox$ may reflect a change in the dominant emission processes in the hot corona. As suggested for XRBs, in this low accretion regime ($\eddfrac \sim 0.0001 - 0.001$), the UV emission can become stronger relative to the X-ray emission due to a change in emission mechanisms \citep{sobo2011MNRAS.417..280S}. In particular, cyclo-synchrotron emission from a hot inner flow \citep{Narayan1995ApJ...452..710N,Veledina2011ApJ...737L..17V} or contributions from a weak jet \citep{zdz2003MNRAS.342..355Z,Markoff2005ApJ...635.1203M} may become important, thereby enhancing the UV/optical emission relative to the X-rays \citep{Ruan2019ApJ...883...76R}.

At values higher than the critical value $\eddfrac \sim 0.02$, the trend reverses, suggesting a transition in the disk-corona configuration. Now, the optically thick disk likely extends inward toward the innermost stable circular orbit, increasing the UV emission, while the emergence of a radiatively efficient warm corona enhances the reprocessing of disk photons into the soft X-ray band. The increased supply of seed photons leads to more efficient cooling of the hot corona, resulting in X-ray spectral softening and an increase in $\aox$. In this regime, both the thermal disk and the warm corona contribute significantly to the energy output. In XRBs, this is analogous to the high/soft state, where the disk emission becomes dominant and the coronal component is strongly suppressed \citep{Dunn2010MNRAS.403...61D,belloni2016ASSL..440...61B}.




In contrast to the standard $\aox$ and $\aoxtwo$ trends shown in Fig.~\ref{fig:alpha_ox1}, the $\aoxone$-$\eddfrac$ relation exhibits a divergent behavior, showing no clear break point. The shared variability between the thermal disk and the warm corona suggests a tighter radiative coupling between these components, causing the observed continuous declining trend for the $\aoxone-\eddfrac$ relation. This notion agrees with the previously proposed geometries for the inner accretion flow, where the warm corona is situated on top of a passive accretion disk \citep{POP2018A&A...611A..59P,2024palit}. Additionally, magnetic energy dissipation in accretion disks has been shown to naturally produce such optically-thick warm layers in both XRBs and AGNs\citep{gronk2020A&A...633A..35G,gronk2023A&A...675A.198G}. This configuration enables efficient Compton up-scattering of UV seed photons, enhancing the soft X-ray
excess and contributing significantly to the ionizing radiation field. Meanwhile, a warm corona was already present at $\eddfrac \sim 0.01$, although the SXE flux remained weak \citep{2025Lawther}. Only in the most recent epochs  -- since early 2024 --  did the SXE flux strengthen significantly, possibly indicating that the warm corona became dissipative as the source transitioned into a higher accretion state \citep{2025Palit}. 

Using two independent tracers of accretion power, $\log(\lambda L_{2500}/L_{\rm Edd})$ and $\log(L_{\mathrm{X}}/L_{\rm Edd})$, we found consistent break points at $ -2.46 \pm 0.14$ and $-2.41 \pm 0.01$, respectively. Comparable break point values using $\log(\lambda L_{2500}/L_{\rm Edd})$ as a tracer have been reported in other well-studied CLAGNs such as NGC 2617 and ZTF18aajupnt \citep{2019arXivRuan}. In a recent study of SXE evolution on a sample of CLAGNs, a turnover at $\log \eddfrac = -2.47 \pm 0.09$ in the hot corona photon index ($\Gamma$) vs $\eddfrac$ plane was reported 
\citep{Jana2026arXiv260107337J}, consistent with earlier findings in both high- and low-luminosity AGNs \citep{she2018ApJ...859..152S}. Interestingly, as demonstrated in Fig.~\ref{fig:LX_uv}, below the break point at $\log(\lambda L_{2500}/L_{\rm Edd}) \sim -2.44$, the X-ray luminosity scales more rapidly than the UV luminosity, whereas above it the relation becomes shallower, suggesting that the relative scaling between the UV-emitting disk and the X-ray-emitting corona changes across the transition.
Additionally, it is worth noting that simulated values of $\aox$ for SMBHs (green squares in Fig.~\ref{fig:grand_aox}), 
versus  $\log(\lambda L_{2500}/L_{\rm Edd})$ display a break at $\sim -3.5$ \citep[Fig.~11 in][]{2019arXivRuan}, distinct from the break point around $-2.4$ in our work. This discrepancy may reflect limitations in the assumed models for scaling canonical XRB accretion states to CLAGNs.

\subsection{Phenomenological accretion phases of Mkn~590}
\label{sec:HID_conc}
We have demonstrated in this paper that the recent evolution of Mkn~590 can be divided into a sequence of broad phenomenological phases, each characterized by distinct combinations of luminosity, spectral hardness, and UV-to-X-ray balance. 
These phases are defined empirically from the monitoring data and we do not attempt to find analogies between each individual state in Mkn~590 and those in XRBs. Nevertheless, the rapid rise in Eddington fraction during mid-2024, occurring within $\sim$8 months, together with the non-monotonic paths in the diagnostic plots presented in Sec.~\ref{sec:hids}, is qualitatively reminiscent of state transitions seen in XRBs. We define the following accretion phases of Mkn~590:


A -- Dec. 2013 to Mar. 2017 -- Faint phase. 
Characterized by a low accretion rate, $\eddfrac \sim 0.003-0.004$, with relatively hard X-ray spectrum and overall low luminosity.
Observations from \textit{Suzaku} in 2011, and 
 \textit{Chandra} in 2014, reveal a flat coronal X-ray continuum, with $\Gamma= 1.67\pm0.01$  \citep{2012rivers}, and $\Gamma= 1.6\pm0.1$  \citep{2018Mathur}, respectively.  
 Such low values are often observed for low accretion rates in both XRBs and AGN, where radiatively inefficient inner flows or truncated-disk geometries are invoked. 
 
B -- Mar. 2017 to Oct. 2023 -- Flaring phase. The source displayed strong fluctuations in $\eddfrac$ accompanied by multi-wavelength variability on the timescales $\sim 100$ days, consistent with the thermal timescales operating in the inner disk \citep{lawther2023MNRAS}. The UV and X-ray continuum rose by factors of a few since the Faint phase (A), also marked by re-appearance of the broad Balmer lines \citep{Raimundo2017,lawther2023MNRAS}. The strongly correlated X-ray/UV variability reported by \citet{lawther2023MNRAS} during this phase is shown to evolve as an  irregular or loop-like -- like trajectory on the diagnostic plots (Fig.~\ref{fig:HID}) and momentarily (when $\eddfrac < 0.02$) showcases a harder-when-brighter trend in the $\eddfrac - HR$ diagram (Fig.~\ref{fig:HID}, middle panel) expected during the canonical hard state in XRBs.

C -- Oct. 2023 to Mar. 2024 -- Transition phase. 
As discussed in Sec.~\ref{sec:hids}, the UV flux dropped sharply in comparison to X-rays, while $HR$ softened slightly. 
This phase coincides with the onset of a relative dominance of soft X-ray emission, as inferred from the intermediate values of $HR$ and the subsequent rapid increase in the soft X-ray excess flux. In the epochs immediately following this transition, the SXE flux rises by nearly two orders of magnitude corresponding to observations from Flaring phase (B) (top panel of Fig.~\ref{fig:HID}), pointing to a substantial change in the geometry or energetics of the warm corona. This behavior is consistent with scenarios in which the warm corona becomes more dissipative or increases its covering fraction over the inner disk, thereby intercepting a larger fraction of UV photons. Similar behavior has been proposed for the luminous phases of the CLAGNs Mkn~1018 \citep{saha2025A&A...699A.205S} and NGC~1566 \citep{tripathi2022ApJ...930..117T}. Such enhanced Comptonization can naturally explain the observed drop in UV emission in Mkn~590. The duration of this transition, from B8$\rightarrow$D1 through C is $\sim14$ months, and it is significantly shorter than the decade long changing look cycle but longer than the month scale variability seen during the Flaring phase. This intermediate timescale likely reflects a rapid reconfiguration of the inner disk/corona system, potentially triggered by disk instabilities \citep{Noda2018MNRAS.480.3898N}.


D -- July 2024 to Aug. 2025 -- Bright phase. 
This represents the current state of the source, marked by a sharp rise in $\eddfrac$, where the overall emission is dominated by soft X-rays and a relative dominance of UV emission over the hard X-rays (yellow regions in the bottom panels of Fig.~\ref{fig:HID}). This drives the renewed ionization of the BLR and the recent strengthening of the broad Balmer lines \citep{2025Palit}.  It constitutes the right arm of the \lq V\rq- trend and exhibits softer-when-brighter trend. 
XRBs commonly undergo such hard-to-soft state transitions via outbursts, where the $\eddfrac$ rises by a few orders of magnitude. However, the switch from  harder-when-brighter to softer-when-brighter in XRBs occur during the canonical hard state and the softer-when-brighter continues well into the canonical hard-intermediate state  \citep{skipper2016MNRAS.458.1696S}. Thus, most likely, even in its current bright phase of evolution, Mkn~590 resides in the hard-intermediate branch of canonical XRB states \citep{connoly2016MNRAS.459.3963C,Ruan2019ApJ...883...76R}.

Continued monitoring will be important to determine whether the source evolves toward even higher accretion rates and a further softening of $\aox$, entering the analogous regime of canonical high/soft state in XRBs, or whether it remains confined to the current bright phase. 

 Overall, the coexistence of month-, year-, and decade-scale variability likely reflects different physical processes. Mkn~590 is a recurring CLAGN, having showcased three changing look events in $\sim 50$ years (Fig.~\ref{fig:history}). The gradual dimming starting in the mid-1990s lasted for $\approx 20$ years until 2015 (A in Fig.~\ref{fig:HID}).
 
 The total time of phase transition, spanning between the end stages of fading (A) and  complete reawakening (D2), is $\sim 10$ years. 
The timescales constrained from sample studies of CLAGNs indicate a  \lq turn-off\rq\, to \lq turn-on\rq\, time difference of 5--20 years \citep{wang_timescales2025ApJ...981..129W,Dong2025arXiv251018445D,2024Panda, jana2025A&A...693A..35J}, with the \lq turn-off\rq\, phase lasting longer than \lq turn-on\rq\, phase \citep{DESIGuo_2025}, which is consistent with our results. Such timescales are broadly consistent with models invoking propagating thermal fronts in radiation- or magnetically supported accretion disks \citep[e.g.,][]{Noda2018MNRAS.480.3898N,2018stern, 2018Ross}, which can increase the disk scale height and drive rapid reconfiguration of the inner disk-corona structure. 

\subsection{Long term radio/X-ray variability}
\label{sec:FP_conc}
Here, we interpret the long-term X-ray and radio variability of Mkn~590 over the past five decades (Fig.~\ref{fig:history}) As noted by \citet{2016koay}, assessing intrinsic variability in the GHz radio band is
complicated by the use of instruments with differing beam sizes.
To minimize these uncertainties, we select seven epochs of quasi-simultaneous X-ray and radio observations of Mkn~590 to access the long-term radio-X-ray connection. 
Furthermore, we  compared our results with another prototypical CLAGN -- Mkn~1018, for which only five such epochs have been reported \citep{lyu2021MNRAS.506.4188L}. For both sources, the radio-to-X-ray luminosity ratio decreases with increasing Eddington fraction,
suggesting that this behavior may be a common feature of CLAGNs. Close to a few percent of $L_{\rm bol}$, both Mkn~590 and Mkn~1018 have $\log(\nu L_{\rm 5\,GHz}/L_{\rm 2-10\,keV}) \lesssim -4.5$, comparable to values observed in low-luminosity AGNs where compact jets can contribute significantly to the nuclear radio emission \citep{terashima2003ApJ...583..145T}. This interpretation is supported by the VLBA detection of a parsec-scale radio jet in Mkn~590 in 2015 \citep{2021Yang}. At higher accretion rates, the lower radio/X-ray ratio observed in both the sources may reflect the suppression of a steady jet. 
The decline of the radio-to-X-ray luminosity ratio with increasing $\eddfrac$ is qualitatively reminiscent of XRBs, in which low/hard states (typically at $\eddfrac \ll 10^{-4}$) are associated with compact steady jets, while high/soft states ($\eddfrac \gtrsim 10^{-1}$) show reduced radio jet activity and stronger disk emission \citep{2004fender, Maccarone2003MNRAS.345L..19M}.

A notable feature in the historical light curve of Mkn~590 is the radio re-brightening observed in the mid-1990s following the rise in X-ray luminosity that began in the late 1980s (Fig.~\ref{fig:history}). If the source underwent an accretion state transition during this interval, the delayed radio brightening may indicate a transient mass ejection that could be triggered by the shocked outflow in the innermost, Comptonized region,
and resulting in larger variability in the GHz band, arising from coronal activity $\log( \nu L_{\rm 5 GHz}/L_{\rm 2-10 keV})$, $\sim - 5$,  during the bright state \citep{gudel1993ApJ...415..236G,2016koay}.
This interpretation is further supported by observations of XRBs, in which transition from hard to soft spectral states 
is frequently associated with the suppression 
 of the compact Comptonizing corona and the subsequent launch of discrete, optically thin, relativistic ejecta detectable in the radio band. This behavior is especially well-documented in the case of GRS 1915+105 \citep{Vadawale2003ApJ...597.1023V, 2004fender}. As shown in the \textit{Swift} X-ray/UV light curves, a short period, pre-flare dip was detected for Mkn~590 (also see Fig.~\ref{fig:HID}, top panel, B8). In GRS 1915+105, similar X-ray dips have been temporally linked to the onset of infrared and radio flares, and interpreted as signatures of matter being expelled from the inner disk \citep{pooley1997MNRAS.292..925P,mirabel1998A&A...330L...9M}. In another AGN with repeating changing look events, namely, NGC 1566, high resolution X-ray observations revealed the appearance of a $\sim 500$ km s$^{-1}$ outflow after a changing look event \citep{parker2019MNRAS.483L..88P}. Hence, if Mkn~590 is currently undergoing a similar accretion cycle as suggested by the moderate rise in radio flux density detected in early 2025 \citep{2025palit_borkar}, a comparable delayed radio flare may be imminent, similar to observations of the CLAGN 1ES 1927+654 \citep{meyer2025ApJ...979L...2M}. This reinforces the need for sustained, high-cadence monitoring of Mkn~590 in both radio and X-ray bands over the coming years, which could offer a rare opportunity to directly capture jet reactivation linked to state evolution in this CLAGN.


\subsection{On systematic biases in bolometric corrections for CLAGN}
\label{subsec:Lbol_clagn}
We show in Appendix~\ref{appendixC} that care must be exercised when determining the bolometric luminosity of AGN with dramatic variability amplitudes, such as CLAGN. Existing standard bolometric correction schemes, although carefully determined based on large samples of high-quality data, inherently assume the intrinsic SED is the same for all AGN and that the SED does not change during source variability. In addition, the bolometric corrections are based on type 1 AGN with significant X-ray detections, which potentially selects X-ray brighter AGN, and the sample does not include many, if any, AGN with dramatic source variability as we see for Mkn~590.

Comparing with the bolometric luminosity measured by integrating over the Swift-observed SED, we find that standard bolometric correction factors tend to overestimate the bolometric luminosity for Mkn 590 by factors of 2 to 3 (0.35 -- 0.5 dex; Fig.~\ref{fig:compare}). The offsets are significant at the 4.5--5 $\sigma$ level. If used to represent the Eddington luminosity ratio in the $\aox-\eddfrac$ relation, the break point would be significantly overestimated. This emphasizes that for better constraints on the underlying physics of AGN, directly measured $L_{\rm bol}$ values are preferred when the relevant observations are available.

\section{\bf Conclusions}
\label{sec:summary}
Our main conclusions are as follows:
\begin{enumerate}
\item We detect a pronounced \lq V\rq-shaped $\alpha_{\rm ox}-\eddfrac$ relation with a break at $\eddfrac = 0.021 \pm 0.008$, indicating a transition in the disk-corona structure analogous to spectral state transitions observed for XRBs. The behavior is consistent with a change from a truncated disk with dominant hot corona at $\eddfrac \lesssim 0.01$ to an inward-extending disk with a prominent warm corona at $0.01 \lesssim \eddfrac \lesssim 0.1$ \citep{Esin1997ApJ...489..865E, 2007Done,POP2018A&A...611A..59P}. Furthermore, a consistent break is observed at $\eddfrac \sim 0.004$ using both UV and X-ray Eddington ratio tracers  (Sec.~\ref{sec:alfy}, Sec.~\ref{sec:aox_conc}, Appendix~\ref{appendixC.2}).
\item The absence of a break in the  $\aoxone-\eddfrac$ relation (Sec.~\ref{sec:alfy}) indicates a tight radiative coupling between the thermal disk and the warm corona, consistent with a geometry in which the warm corona forms an optically thick layer above the disk (Sec.~\ref{sec:aox_conc}).

\item The source evolves through distinct phenomenological accretion phases- faint, flaring, transitional, and bright phases -- lasting $\sim 10$ yr.  This timescale is shorter than classical viscous timescales but broadly consistent with propagating thermal fronts in the accretion disk (Sec.~\ref{sec:hids} and Sec.~\ref{sec:HID_conc}).

\item Comparing Mkn~590 with the CLAGN Mkn~1018, we find that the radio-to-X-ray luminosity ratio decreases with increasing accretion rate in both sources. This trend is qualitatively consistent with XRB-typical behavior, where compact jets dominate at low accretion rates and become suppressed at higher rates, However, Seyferts likely retain residual coronal radio emission. The radio re-brightening in Mkn~590 may suggests a possible transient jet ejection associated with its accretion state transition (Sec.~\ref{sec:fund} and Sec.~\ref{sec:FP_conc}).

\end{enumerate}
\begin{acknowledgments}
We would like to thank the referee for useful comments and suggestions. BP has been fully and AR has been partially supported by Polish National Science Center (NCN) grant No.\ 2021/41/B/ST9/04110. AM acknowledges support from NCN grant 2018/31/G/ST9/03224. MV, DL and GW acknowledge financial support by the Independent Research Fund Denmark via grants DFF-8021-0013 and DFF-3103-00146. Also, this work has benefited from Swift observing programs supported by the Instrument Centre for Danish Astronomy. DL acknowledges support from NASA grant 22-SWIFT22-0029 for this work. AB acknowledges the support of the EU-ARC.CZ Large Research Infrastructure grant project LM2023059 of the Ministry of Education, Youth and Sports of the Czech Republic. MS acknowledges Czech Science Foundation (GA\v{C}R) grant no.\ 26-23342I. KXL acknowledges financial support from the National Natural Science Foundation of China (12573020), and the Young Talent Project of Yunnan Province. BP acknowledges Dr.\ Sandra Raimundo for insightful discussions during the early stages of this work, and Dr.\ Swayamtrupta Panda for kindly providing access to the DESI spectrum.
\end{acknowledgments}

\begin{contribution}

This work is based on more than a decade of public {\it Swift}
 monitoring observations initiated by numerous members of the
 community, in particular M.V.\ and D.L. The core analysis was enabled by data secured by M.V., D.L. and G.W.\, B.P. led the X-ray/UV data reduction, performed the primary analysis, and wrote the manuscript. The investigation of the long-term variability in $\alpha_{\rm ox}$ was independently initiated by B.P. and D.L.\, B.P. A.R. and A.G.M.\ advanced the study to cover accretion state transitions and comparisons with X-ray binaries.  Everyone contributed through scientific discussions, critical feedback, and manuscript input that shaped the interpretation and presentation of the results. The hardness-intensity diagram analysis benefited in particular from contributions by A.R. and A.G.M., while M.V. contributed substantially to the bolometric correction methodology presented in Sec.~7.4 and Appendix~\ref{appendixC}.\, M.V. and D.L. provided the Mkn~590 host galaxy spectrum used in the analysis. J.J.R. provided the archival data essential for Fig.~\ref{fig:grand_aox}.\, G.W. and A.B. provided feedback on the radio analysis. T.S. provided input on the X-ray/UV analysis. A.B. led the reduction and analysis of the GMRT and ASKAP/VAST radio datasets. M.S. contributed to the modeling of the NOT/ALFOSC and DESI spectra. K.-X.L. contributed to an early discussion that was motivated by long-term optical monitoring of the source. 

\end{contribution}

%
\facilities{Swift(XRT and UVOT), GMRT, NOT}

\software{PyXSPEC \citep{pyxspec2021ascl.soft01014G}, PyQSOFIT \citep{pyqsofit2018ascl.soft09008G}, Matplotlib \citep{matplotlib2007CSE.....9...90H}, Numpy \citep{numpy2020Natur.585..357H}, Astropy \citep{astropy2022ApJ...935..167A} }


\appendix

\section{{\bf Details on the historical data}}
\label{appendixA}

\subsection{Our laboratory: Mkn~590}
\label{appendixA.1}
Mkn~590 ($z$ = 0.0264), located in the direction of the constellation Cetus, stands out as one of the most important nearby CLAGN, having exhibited multiple changing look events over the past five decades.  An overview of its long-term activity is shown in Fig.~\ref{fig:history} and Fig. 4 of \citet{2014Denney}.

A compilation of optical, UV, and X-ray observations spanning from 1970 to 2015 reveals at least two major changing look events: first, there was a brightening or  \lq turn-on\rq\ phase between 1973 and 1989. During the early 1970s, optical observations from the Lick Observatory classified Mkn~590 as a Sy\,1.5 within the AGN unified scheme \citep{1993antonucci,1995Urry}, characterized by relatively modest but detectable broad H$\beta$ emission. 
Subsequent observing campaigns, using the 1.8\,m Perkins telescope at Lowell Observatory, indicate the 5100\AA\ continuum luminosity increasing by a factor of $\sim$ 16 through the early 1980s up to the early 1990s, while the broad H$\beta$ component underwent significant strengthening in flux, marking a clear transition to a Sy\,1 activity state \citep{Ferland1990ApJ...363L..21F}. As shown in Fig.~\ref{fig:history} using green and blue markers, this period was also characterized by the increased emission in UV and X-rays as observed by some of the earliest high energy missions -- the \textit{International Ultraviolet Explorer} (\textit{IUE}), the \textit{European X-ray Observatory Satellite} (\textit{EXOSAT}) and the \textit{ROentgen SATellite} (\textit{ROSAT}). The 1450\AA\ UV continuum and 2--10 keV integrated hard X-ray emission increased by factors of $\sim$ 5 and 2 respectively, thus showing a near-simultaneous response across multiple wavebands \citep{turner1989MNRAS.240..833T,voges1999A&A...349..389V}.

Then, from the late 1990s through the 2010s, the source entered a shutdown phase, during which the luminosity across all wavebands underwent a steady decline \citep{2014Denney}. A 2003 SDSS optical spectrum shows only a weak broad H$\beta$ component; by then, the optical continuum had dropped by an order of magnitude. Meanwhile, X-ray observations with \textit{XMM-Newton} and \textit{Chandra} tracked a drop in X-ray flux by a factor of $\sim 10$ from the early 1990s to 2004 \citep{longiniotti2007A&A...470...73L}. By 2014, Mkn~590 had reached a historic low state, with X-ray and optical fluxes reduced by $\sim 2 $ orders of magnitude relative to its peak phase in the 1990s, while the optical spectral type had transitioned to Sy\,1.8-1.9, marking a second changing look transition. Despite this extreme low state, evidence for the persistence of the BLR was found via UV spectroscopy, which shows a continued presence of the \ion{Mg}{2} $\lambda\lambda2796,2803$ emission lines \citep{2018Mathur}, while optical spectroscopy obtained during 2017--2018 with VLT/MUSE and Subaru/HDS reports weak but significant broad Balmer lines \citep{Raimundo2017,mandal2021MNRAS}. These observations suggest that the BLR in Mkn~590 remained largely intact even as the ionizing continuum weakened substantially. 

\citet{lawther2023MNRAS} presents a detailed light curve of Mkn\,590 which exhibits renewed flaring activity in the X-ray, UV, and optical bands on timescales of $\sim$100 days since 2017. It is broadly consistent with the thermal timescales as expected from the standard accretion disk theory. This flaring subsided by the end of 2022, as the source momentarily lowered its UV/X-ray intensity \citep{2025Lawther}. 

This brief low-flux period ended in an abrupt and
sustained spike in broadband fluxes post-2023, marking the second major re-brightening episode in fifty years.
As tracked with \textit{Swift},
the far-UV and X-ray fluxes increased by factors of $\sim 12$ and $\sim 15$, respectively, from late 2023 through early 2025, the largest such increase since $\sim 1990$ 
\citep{2025Lawther,2025Palit}. Timely optical spectroscopic observations taken in Nov.\ 2024 at Siding Spring Observatory confirms the reappearance of the broad H$\beta$ component, with its flux now six times stronger than in 2003, signalling a changing look transition of Mkn~590 back to a Sy 1 activity type \citep{2025Palit}.

As illustrated by the cyan diamonds in Fig.~\ref{fig:history}, the SXE in Mkn~590, visible in the early 2000s, had completely disappeared by 2008--2011 \citep{2012rivers}. SXE flux measurements were obtained by fitting a distinct spectral component in the 0.2--2~keV band.  A weak SXE emission component reappeared in 2014, suggesting the start of the episodic accretion activity  observed in the years following \citep{2018Mathur}. Until 2021, Mkn 590 had not exhibited a full recovery of the SXE, with only a weak component detected \citep{ghosh2022}. Based on {\it XMM-Newton} observations obtained in low states and higher, flaring states during years 2020 - 2024, \citet{2025Lawther} do not find significant evidence that the SXE component had changed \footnote{While \citet{ghosh2022} report a lack of SXE in {\it Swift} data since 2010, \citet{2025Lawther} use simultaneous {\it Swift} XRT and {\it XMM-Newton} observations to show that {\it Swift} XRT is not sufficiently sensitive to detect the weak SXE from Mkn~590.} since 2004. Notably, the changing look event of 2025 did convincingly reveal the emergence of an independent, strong SXE component, likely associated with the warm corona and rising faster than the hard X-ray power-law emission \citep{2025Palit}.

Mkn~590 is also variable in the GHz regime, and has been monitored across three decades \citep{2016koay}.
 The earliest reported radio observation dates back to 1977, when the source was detected with the Westerbork Synthesis Radio Telescope (WSRT) at a flux density of 11~mJy in the 1.4~GHz band\citep{1982wilson}. Over the following two decades, multiple arcsec-resolution observations with the Very Large Array (VLA) spanning frequencies from 1.4 to 8.4~GHz measure a mean flux density of $\sim 4$~mJy. After the turn of the century, the radio emission intensity starts to exhibit a gradual decline through 2015 where fresh VLA measurements recorded a historic low core radio flux density of 3 mJy \citep{2016koay}. This coincided with detection of a faint, radio jet at a flux density of $\sim$1.7~mJy  at 1.6 GHz using high-resolution sub-arcsec resolution measurements by the European VLBI Network (EVN)  \citep{2021Yang}. More recent observations with the Australian Square Kilometer Array Pathfinder (ASKAP) as part of the Variables and Slow Transients (VAST) survey from mid-2022 \citep{sufia2025arXiv250701355B} and the Giant Meterwave Radio Telescope (GMRT) in early 2025 \citep{2025palit_borkar} reports a modest increase in GHz-band radio flux densities, reaching levels of up to $\sim$ 5 mJy, and indicating a upward rising trend, tracking the contemporaneous brightening observed at other wavelengths and approaching flux levels last seen in the early 1980s. Interestingly, the 2015 radio measurements are consistent with the Fundamental Plane of black hole activity for sources in the low/hard state \citep{2003merloni,2016koay}.

\subsection{Historical data collection:}
\label{app:1018}
Here, we detail on the multi-band data gathered across the entire history of the Mkn~590, and summarized in Tab.~\ref{tab:mrk590_continuum}. Owing to the heterogeneous nature of the dataset, measurement uncertainties for the optical, UV, and X-ray fluxes are not uniformly available and are therefore not included. The long-term light curve shown in Fig.~\ref{fig:history} is intended to illustrate the overall multi-wavelength variability trends only. The majority of the optical continuum fluxes are compiled from published studies, where full details of the observing campaigns are described in the cited papers and references therein. The NOT/ALFOSC spectrum was obtained through open calls for observation proposals (70-406; PI: B. Palit). The DESI and NOT/ALFOSC spectra were flux-normalized using the [O III]
$\lambda$5007 narrow emission line following the framework of \citet{vangronoingren1992PASP..104..700V}, adopting as reference the line flux measured in the 2003 SDSS spectrum \citep{2025Palit}, and using the fitting software PYQSOFIT \citep{pyqsofit2018ascl.soft09008G}. Both spectra were presented in \citet{2025palit_borkar} and will be analyzed in detail in a forthcoming study focused on broad-line region modelling.

The reported UV fluxes were obtained using broadband filters indicated in parentheses, and the continuum region corresponds to their respective effective wavelengths. 
Owing to the heterogeneous energy coverage of different X-ray instruments, we report fluxes integrated over their respective observing energy bands. For the purposes of the Fundamental Plane analysis (Sec.~\ref{sec:radio}), the fluxes were converted to a uniform energy range using \texttt{WebPIMMS} tool. Between 2015-2025, we selected 2--4 \textit{Swift}-XRT pointings per year to show the approximate trend in X-rays. The SXE measurements were only possible post-2000 and have been taken from \citet{ghosh2022,2025Lawther,2025Palit}. These fluxes were estimated after fitting a separate spectral component to model the SXE. 

The radio observations vary in both beam size and continuum frequency, introducing non-uniformity in the historical light curve and the radio/X-ray variability. To minimize potential biases, we restrict our sample to observations with arcsec-scale resolution and frequencies below 10~GHz.

\startlongtable
\begin{deluxetable}{ccccc}
\tabletypesize{\footnotesize}
\tablecolumns{5}
\tablecaption{Mkn~590 continuum properties used in this paper.}
\tablehead{
\colhead{Observation Identifier} &
\colhead{Year} &
\colhead{Continuum Region} &
\colhead{Continuum Flux$^{d}$} &
\colhead{Reference$^{e}$}
\\
(1) & (2) & (3) & (4) & (5)
}
\startdata
\cutinhead{Optical Spectra$^{a}$}
Lick IDS        & 1973    & 5100 \AA      & 3.4      & D14 \\
Perkins/OSU IDS & 1983    & 5100 \AA      & 11.0     & D14 \\
RM campaign     & 1989    & 5100 \AA      & 55.0     & P04, D14 \\
RM campaign     & 1993    & 5100 \AA      & 26.0     & P04, D14 \\
RM campaign     & 1996    & 5100 \AA      & 42.0     & P04, D14 \\
SDSS            & 2003    & 5100 \AA      & 1.30     & D14 \\
MDM             & 2006    & 5100 \AA      & 0.28     & D14 \\
LBT MODS1       & 2013    & 5100 \AA      & $<0.014$ & D14 \\
LBT MODS1       & 2013    & 5100 \AA      & $<0.014$ & D14 \\
LBT MODS1       & 2013    & 5100 \AA      & $<0.014$ & D14 \\
KOSMOS          & 2013    & 5100 \AA      & $<0.10$  & D14 \\
MDM             & 2014    & 5100 \AA      & $<0.11$  & D14 \\
MUSE/VLT        & 2017    & 5100 \AA      & 45.0     & R17 \\
Subaru/HDS      & 2018    & 5100 \AA      & 19.0     & M21 \\
DESI            & 2022    & 5100 \AA      & 4.35     & P26 \\
LCO/FLOYD       & 2024    & 5100 \AA      & 75.0     & P25 \\
NOT/ALFOSC      & 2025    & 5100 \AA      & 118.0    & P26 \\
\cutinhead{UV Spectra$^{b}$}
IUE                     & 1982    & 1450 \AA      & 88.5   & D14 \\
IUE                     & 1991    & 1450 \AA      & 388.0  & D14 \\
XMM/OM (UVW2 filter)    & 2002    & 1928 \AA      & 28.0   & G22 \\
XMM/OM (UVW2 filter)    & 2004    & 1928 \AA      & 26.0   & G22 \\
Swift/UVOT (UVW2 filter)& 2008    & 1928 \AA      & 13.05  & $-$ \\
HST/COS                 & 2013    & 1450 \AA      & 3.7    & D14 \\
HST/COS                 & 2013    & 1450 \AA      & 3.7    & D14 \\
HST/COS                 & 2013    & 1450 \AA      & 3.7    & D14 \\
Swift/UVOT (UVW2 filter)& 2015    & 1928 \AA      & 17.54  & $-$ \\
Swift/UVOT (UVW2 filter)& 2017    & 1928 \AA      & 33.68  &$-$\\
Swift/UVOT (UVW2 filter)& 2020    & 1928 \AA      & 28.90  & $-$ \\
Swift/UVOT (UVW2 filter)& 2022    & 1928 \AA      & 14.45  & $-$ \\
Swift/UVOT (UVW2 filter)& 2024    & 1928 \AA      & 24.01  & $-$ \\
Swift/UVOT (UVW2 filter)& 2025/03 & 1928 \AA      & 175.20 & $-$ \\
Swift/UVOT (UVW2 filter)& 2025/05 & 1928 \AA      & 275.00 & $-$ \\
Swift/UVOT (UVW2 filter)& 2025/09 & 1928 \AA      & 300.00 & $-$ \\
\cutinhead{X-ray Spectra$^{c}$}
Einstein (HEAO-2) & 1979    & 0.4--4 keV    & 13.9  & $\dagger$ \\
Einstein (HEAO-2) & 1979    & 0.4--4 keV    & 10.0  & $\dagger$ \\
EXOSAT            & 1984    & 2--10 keV     & 27.0  & D14 \\
RASS              & 1991    & 0.1--2.4 keV  & 46.3  & D14 \\
ROSAT             & 1995    & 0.1--2.4 keV  & 3.47  & $\dagger$ \\
ROSAT             & 1996    & 0.1--2.4 keV  & 11.0  & $\dagger$ \\
XMM               & 2004    & 0.2--2 keV    & 3.31  & D14 \\
XMM               & 2004    & 2--10 keV     & 6.95  & D14 \\
Chandra           & 2004    & 0.5--10 keV   & 11.9  & D14 \\
Swift/XRT         & 2008    & 2--10 keV     & 3.6   & $-$ \\
Suzaku/XIS+PIN    & 2011    & 2--10 keV     & 6.8   & D14 \\
Chandra           & 2013    & 0.5--10 keV   & 1.3   & D14 \\
Swift/XRT         & 2016/02 & 2--10 keV     & 4.9   & $-$ \\
Swift/XRT         & 2016/12 & 2--10 keV     & 5.0   & $-$ \\
Swift/XRT         & 2018    & 2--10 keV     & 14.8  & $-$ \\
Swift/XRT         & 2019    & 2--10 keV     & 21.8  & $-$ \\
Swift/XRT         & 2020/01 & 2--10 keV     & 20.9  & $-$ \\
Swift/XRT         & 2020/07 & 2--10 keV     & 10.8  & $-$ \\
Swift/XRT         & 2021/01 & 2--10 keV     & 10.9  & $-$ \\
Swift/XRT         & 2021/08 & 2--10 keV     & 18.8  & $-$ \\
Swift/XRT         & 2022/01 & 2--10 keV     & 3.8   & $-$ \\
Swift/XRT         & 2022/07 & 2--10 keV     & 9.9   & $-$ \\
Swift/XRT         & 2022/08 & 2--10 keV     & 7.7   & $-$ \\
Swift/XRT         & 2022/09 & 2--10 keV     & 9.9   & $-$ \\
Swift/XRT         & 2022/10 & 2--10 keV     & 6.1   & $-$ \\
Swift/XRT         & 2022/11 & 2--10 keV     & 6.2   & $-$ \\
Swift/XRT         & 2022/12 & 2--10 keV     & 5.2   & $-$ \\
Swift/XRT         & 2023/01 & 2--10 keV     & 5.0   & $-$ \\
Swift/XRT         & 2023/09 & 2--10 keV     & 5.5   & $-$ \\
Swift/XRT         & 2024/01 & 2--10 keV     & 21.5  & $-$ \\
Swift/XRT         & 2024/07 & 2--10 keV     & 20.5  & $-$ \\
Swift/XRT         & 2024/09 & 2--10 keV     & 28.2  & $-$ \\
Swift/XRT         & 2024/12 & 2--10 keV     & 36.6  & $-$ \\
Swift/XRT         & 2025/01 & 2--10 keV     & 27.1  & $-$ \\
Swift/XRT         & 2025/03 & 2--10 keV     & 19.2 & $-$ \\
Swift/XRT         & 2025/05 & 2--10 keV     & 44.0  & $-$ \\
Swift/XRT         & 2025/09 & 2--10 keV     & 39.3 & $-$ \\
\cutinhead{Radio}
WRST   & 1977 & 1.4 GHz ($<$ 13'') & $11.0 \pm 2.0$ & K16 \\
VLA-A  & 1984 & 1.4 GHz (1.95'' $\times$ 1.35'') & $4.86 \pm 0.59$ & K16 \\
VLA-A  & 1991 & 8.4 GHz (0.38'' $\times$ 0.38'') & $3.67 \pm 0.14$ & K16 \\
MERLIN & 1995 & 4.9 GHz (0.33'' $\times$ 0.25'') & $4.23 \pm 0.31$ & K16 \\
VLA-A  & 1998 & 8.4 GHz (0.33'' $\times$ 0.23'') & $3.56 \pm 0.10$ & K16 \\
VLA-B  & 2002 & 1.4 GHz (6.4'' $\times$ 5.4'') & $9.90 \pm 0.10$ & K16 \\
VLA-A  & 2008 & 1.4 GHz (2.35'' $\times$ 2.05'') & $4.35 \pm 0.10$ & K16 \\
VLA    & 2015 & 1.4 GHz (1.99'' $\times$ 1.46'') & $3.39 \pm 0.11$ & K16 \\
VAST   & 2021 & 1.4 GHz (8.89'' $\times$ 7.74'') & 11.93 & B25 \\
GMRT   & 2025 & 1.4 GHz ( 2.39'' $\times$ 2.23'') & $7.43 \pm 0.30$ & P26 \\
\cutinhead{Soft X-ray excess$^{c}$}
XMM-Newton & 2002 & 0.3--2 keV & 4.27    & G22 \\
XMM-Newton & 2004 & 0.3--2 keV & 3.72    &G22 \\
Suzaku     & 2011 & 0.3--2 keV & $<1.6$  & G22 \\
Chandra    & 2014 & 0.3--2 keV & $<3.8$  & M18 \\
Swift/XRT  & 2016 & 0.3--2 keV & $<0.65$ & G22 \\
Swift/XRT  & 2018 & 0.3--2 keV & $<0.55$ & G22 \\
Swift/XRT  & 2020 & 0.3--2 keV & $<1.97$ & G22 \\
XMM-Newton & 2020 & 0.3--2 keV & $7.1$  & L25\\
XMM-Newton & 2021 & 0.3--2 keV & $8.4$  & L25\\
XMM-Newton & 2021 & 0.3--2 keV & $14.1$  & L25\\
Swift/XRT  & 2021 & 0.3--2 keV & $<0.79$ & G22 \\
XMM-Newton & 2022 & 0.3--2 keV & $2.8$  & L25\\
XMM-Newton & 2022 & 0.3--2 keV & $6.1$  & L25\\
XMM-Newton & 2023 & 0.3--2 keV & $5.1$  & L25\\
XMM-Newton & 2024 & 0.3--2 keV & $12.2$  & L25\\
Swift/XRT  & 2024 & 0.3--2 keV & 30.29   & P26\enddata
\tablecomments{
$^{a}$ Optical flux densities derived from spectral fitting as described in \citet{2014Denney}. \\
$^{b}$ UV measurements from IUE and HST/COS are taken from \citet{2014Denney} which were derived from spectral fitting. The specific observations analyzed in \citet{2014Denney} can be accessed via \dataset[DOI: 10.17909/rn31-bb36]{https://doi.org/10.17909/rn31-bb36}. XMM/OM and Swift/UVOT measurements correspond to the central wavelength of the UVW2 filter.  The specific observations by Chandra X-ray Observatory reported here are contained in the Chandra Data Collection ~\dataset[DOI: 10.17909/6tew-qd41]{https://doi.org/10.17909/6tew-qd41}. \\
$^{c}$ Integrated fluxes in the quoted X-ray energy range. \\
$^{d}$ UV/optical fluxes are in units of $10^{-16}\,\mathrm{erg\,s^{-1}\,cm^{-2}\,\AA^{-1}}$; X-ray fluxes are in $10^{-12}\,\mathrm{erg\,s^{-1}\,cm^{-2}}$; radio flux densities are in mJy. Optical, UV and radio continuum fluxes are given at rest-frame wavelength. \\
$^{e}$ $\dagger$ denotes data retrieved from the XMM--Newton upper limit server. $-$ denotes measurements derived in this work.\\
\medskip
\noindent{\it References:}
D14: \citet{2014Denney};
P04: \citet{2004peterson};
R17: \citet{Raimundo2017};
M21: \citet{2021Mondal};
G22: \citet{ghosh2022};
M18: \citet{2018Mathur};
K16: \citet{2016koay};
B25: \citet{sufia2025arXiv250701355B};
P25: \citet{2025Palit};
L25: \citet{2025Lawther};
P26: \citet{2025palit_borkar}.
}
\label{tab:mrk590_continuum}
\end{deluxetable}

\begin{table}[h]
     \caption{The Mkn~1018 radio/X-ray data used in this paper are taken from \citet{lyu2021MNRAS.506.4188L}. The last column includes Eddington ratio estimated from X-ray luminosity ($L_{\rm 2-10 kev}$) following the \citet{Duras2020A&A...636A..73D} prescription.}
         \centering

    \begin{tabular}{|cccc|}
    \hline
         $\log {\nu L_{\rm 5\,GHz}}$& $\log {L_{\rm 2-10\,keV}}$ &$\log\left(\frac{\nu L_{\rm 5\,GHz}}{ L_{\rm 2\text{-}10\,keV}}\right)$ &$\eddfrac$ \\
         \hline
         38.63 & 43.57 & -4.94 & -1.14\\
         38.69& 43.47 & -4.78 &-1.25 \\
         38.68& 42.67 & -3.99 &-2.08 \\
         38.58&  42.96& -4.38 &-1.78 \\
         38.56& 42.89 & -4.33 & -1.85\\
         \hline
    \end{tabular}
   
    \label{tab:lyu}
\end{table}

\section{Bolometric luminosity}
\label{appendixC}
\subsection{Methods of estimating bolometric luminosity}
\label{appendixC.1}
Here we discuss the different methods typically used for estimating the bolometric luminosity and in the subsequent section, we discuss its impact on the $\aox-\eddfrac$ relation. Two of the most widely used methods rely on empirical correlations: one uses the $\aox-L_{\rm 2500\AA}$ linear relation and the other estimates bolometric luminosity from the 2--10 keV integrated X-ray luminosity, both drawn from large samples of bright Seyferts. In the first method, a linear relationship between $L_{\rm 2 keV}$  and  $L_{\rm 2500 \AA}$ is constrained after SED modeling of 545 un-obscured, radio-quiet Type 1 AGNs from the XMM-COSMOS survey, spanning a $z \sim 0.04-4.25$ and four orders of magnitude in $L_{\rm bol}$. From Eq.~11 of \citep{lusso2010A&A...512A..34L}, the derived relation between L$_{\rm bol}$ and $\aox$ is given as: 
\begin{equation}
\log(L_{\mathrm{bol}}) = \log(L_{2-10\,\mathrm{keV}}) + 1.561
 - 1.853\,\aox + 1.226\,\aox^{2} \, .
 \label{eq:lusso}
\end{equation}
\\
In the second method \citep{Duras2020A&A...636A..73D}, the bolometric correction factor 
$K_{\rm bol}$ is based solely on the 2--10 keV X-ray luminosity ($L_X$), derived empirically for a sample of 1000 bright, X-ray selected Type 1 and Type 2 AGNs. Then, bolometric luminosity is estimated as $L_{\rm bol} = K_{\rm bol} \times L_{\rm X}$. Their final sample spans $z \sim 0 - 4$ and four orders of magnitude in $L_{\rm bol}$. We applied the Eq.~3 from \citet{Duras2020A&A...636A..73D} :
\begin{equation}
K_{\rm bol} (L_{\rm X}) = a \left[ 1 + \left( \frac{\log\left( L_{X} / L_{\odot} \right)}{b} \right)^c \, \right],
 \label{eq:duras}
\end{equation}
where the values of a, b, c are determined empirically as 15.33 $\pm$ 0.06, 11.48 $\pm$ 0.01 and 16.20 $\pm$ 0.16, respectively and $L_{\odot}$ is the Solar luminosity equalled $3.83 \times 10^{33}$~erg\,s$^{-1}$. The final expression for $L_{\rm bol}$ along with its propagated uncertainty is :
\begin{align}
\left(
\frac{\sigma_{L_{\rm bol}}}{L_{\rm bol}}
\right)^2
=
&\,
\left(
\frac{\sigma_a}{a}
\right)^2
+
\left(
\frac{c y}{b(1+y)}
\right)^2
\sigma_b^2
+
\left(
\frac{y}{1+y}
\ln\left(\frac{x}{b}\right)
\right)^2
\sigma_c^2
\nonumber \\
&
+
\left(
\frac{c y}{x(1+y)}
\frac{1}{\ln 10}
\right)^2
\left(
\frac{\sigma_{L_X}}{L_X}
\right)^2
+
\left(
\frac{\sigma_{M_{\rm BH}}}{M_{\rm BH}}
\right)^2,
\end{align}


where,
\begin{equation}
x \equiv \log\left(\frac{L_X}{L_\odot}\right),
\qquad
y \equiv \left(\frac{x}{b}\right)^c,
\qquad
K_{\rm bol} = a(1+y).
\end{equation}
\vspace{2mm}

In our work, we employ the SED integration method based on straightforward spectral fitting made with XSPEC software, and integration of the whole flux over the observed wave band.
For each epoch with contemporaneous Swift XRT and UVOT coverage, we constructed a broadband SED and modeled it in three components. 

First, to estimate the integrated UVOT flux at each epoch, we modeled the host-subtracted UVOT SED with a powerlaw in wavelength space, $F_{\lambda} = A\,\lambda^{\alpha}$,
fitted in log--log space using \texttt{weighted linear regression}, where the uncertainty in $\ln F_{\lambda}$ was taken as
$\sigma_{\ln F} = \sigma_{F} / F $.
The best-fit model was then analytically integrated over the wavelength interval spanned by the available UVOT bands for that epoch, $[\lambda_{\min}, \lambda_{\max}]$, to obtain the integrated flux, $F_{\rm int} = \int_{\lambda_{\min}}^{\lambda_{\max}} F_{\lambda}\,{\rm d}\lambda$. Uncertainties were propagated via Monte Carlo resampling: for each epoch, we generated 5000 realizations by drawing flux densities at each wavelength from Gaussian distributions defined by their measured uncertainties, refitted the powerlaw model, and reintegrated it over the same wavelength range. We use the median integrated flux and a $1\sigma$ uncertainty derived from the 16th-84th percentile range of the resulting Monte Carlo distribution. 

Second, a powerlaw model was used to interpolate between the UVW2 band and 0.3 keV, providing an estimate of the extreme-UV contribution. The resulting spectral component was then integrated over the relevant energy range, and the corresponding flux and associated uncertainties are derived following the same procedure described above. Finally, the 0.3-10 keV X-ray spectrum was modeled with an absorbed powerlaw using XSPEC, incorporating Galactic absorption using the \texttt{Tbabs} model \citep{wilms2000ApJ...542..914W}. We adopted a total hydrogen column density of $N_{\rm H} = 2.77 \times 10^{20}\, \mathrm{cm^{-2}}$ \citep{ghosh2022}. No additional intrinsic absorption component was statistically required, supporting the classification of Mkn~590 as a \lq bare\rq\, AGN \citep{2025Lawther}. The best fit model was extrapolated to 50 keV and using the \texttt{flux} command, the integrated X-ray flux was derived. All three fluxes have been added with errors in quadrature and converted to luminosity which corresponds to $L_{\rm bol}$. 

\subsection{The impact of the $L_{\rm bol}$ determination on the $\aox-\eddfrac$ relation}
\label{appendixC.2}
\begin{figure}[h!]
    \centering
    \includegraphics[width=0.98\linewidth]{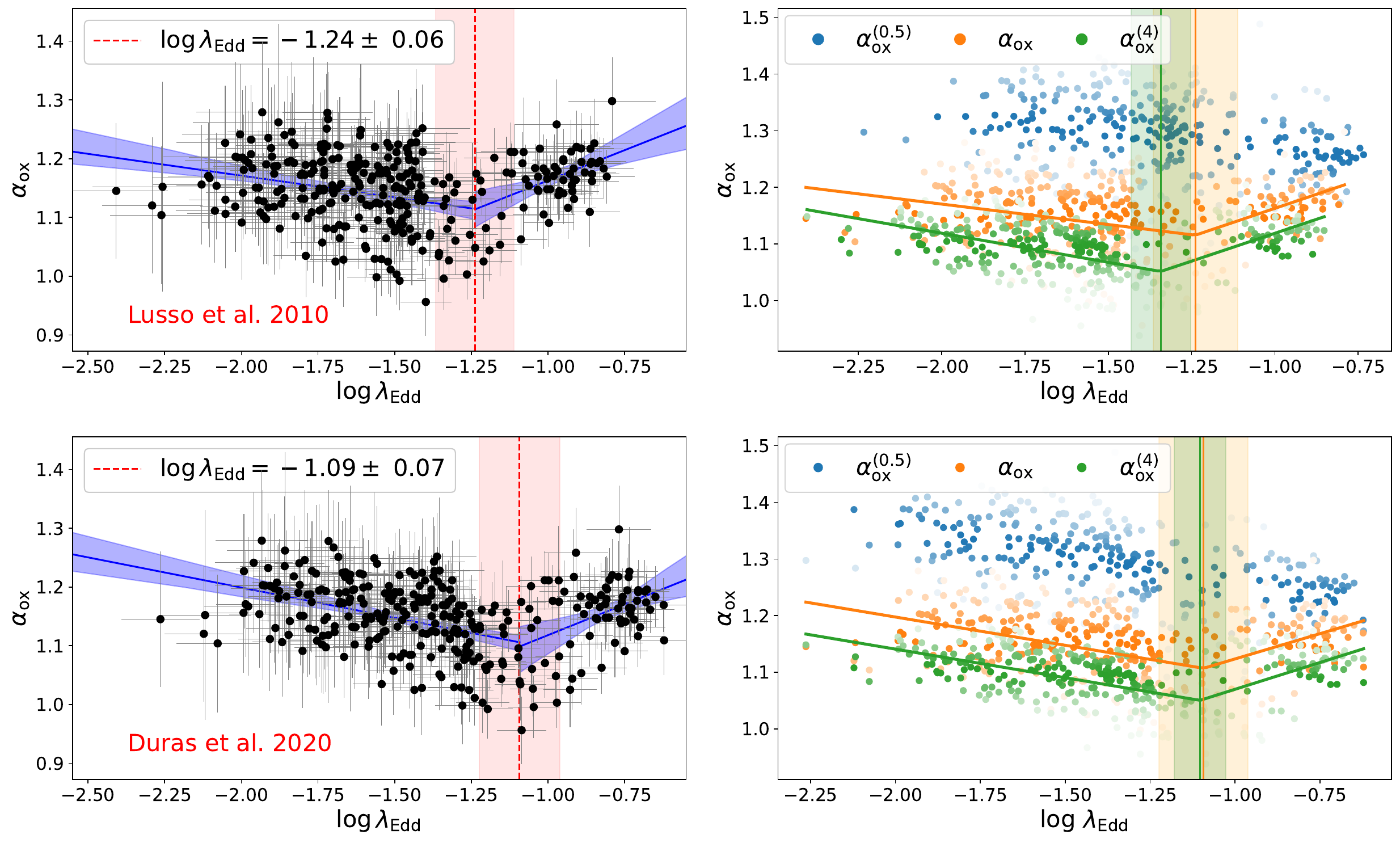}
    \caption{From left to right, this shows the $\aox-\eddfrac$ relation for two distinct methods of bolometric correction factor as discussed in Appendix~\ref{appendixC.1}. Statistical tools used remain the same as discussed in Sec.~\ref{sec:alfy} and Appendix~\ref{appendixD}. }
    \label{fig:lusso_duras}
\end{figure}
As shown in the left panel of Fig.~\ref{fig:lusso_duras}, the first two prescriptions mentioned in Appendix~\ref{appendixC.1} result in different values of the break point, at $\log \eddfrac= -1.23 \pm 0.06$ and $\log \eddfrac= -1.09 \pm 0.06$ respectively. While these breakpoints are mutually consistent within their 1-sigma errors, they both are significantly higher than that obtained when using the SED integration method for the $\eddfrac$ estimate, as can be seen in Fig.~\ref{fig:alpha_ox1} in the main text. The impact of adopting these different $L_{\rm bol}$ estimates on the non-standard $\aox$ definitions introduced in Sec.~\ref{sec:param} is shown in the right panel of Fig.~\ref{fig:lusso_duras}. Thus, the inferred break point(s) is strongly method dependent.

\begin{figure}[h]
    \centering
    \includegraphics[width=0.98\linewidth]{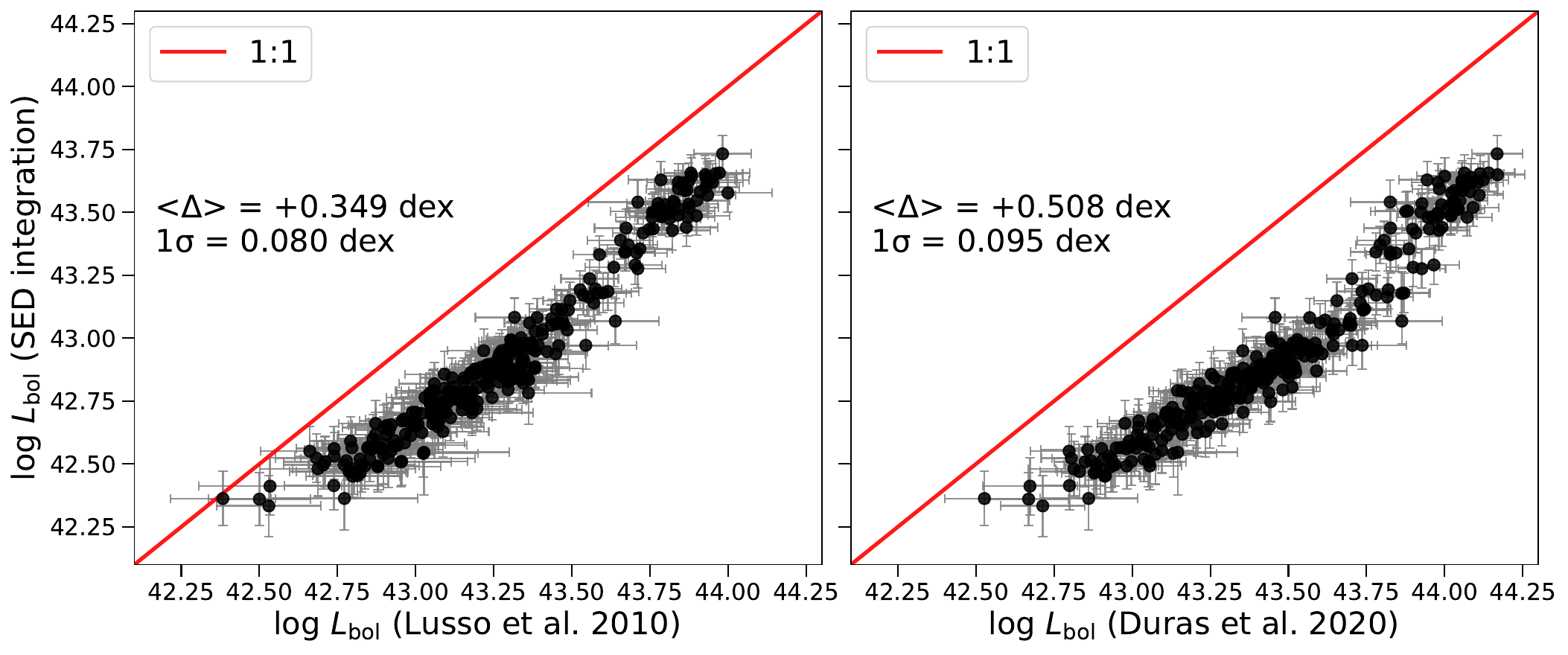}
    \caption{Comparison between bolometric luminosities derived from SED integration method against the two alternate prescriptions described in Sec.~\ref{appendixC.1}. The mean offset factor $<\Delta>$ and its 1$\sigma$ scatter are explicitly annotated.}
    \label{fig:compare}
\end{figure}
 The systematic differences between the SED-integrated luminosities and the two alternative prescriptions are quantified against a 1:1 relation in Fig.~\ref{fig:compare}. Relative to the SED-based values, the \citep{lusso2010A&A...512A..34L} and \citep{Duras2020A&A...636A..73D} methods overestimate $\eddfrac$ by a mean offset factor ($<\Delta>$) of approximately 2.2 and 3.2 (in linear scaling), respectively. These average offsets are statistically significant at the 4.5$\sigma$ - 5$\sigma$ level, as judged by the statistical spread (1 $\sigma$ = 0.08 dex and 0.095 dex, respectively) in this offset.  These overestimated $L_{\rm bol}$ values explain the upward shift in the inferred break point when those prescriptions are adopted.

The fact that the data points trace a curved path in these diagrams show that the SEDs (that of Mkn~590, the population average, or both) are not constant as a function of luminosity.  This comparison emphasizes that dramatically variable AGN, like Mkn 590, do not behave quite like population-typical Type-1 AGN and therefore it is important to measure $L_{\rm bol}$ directly from the observed data when available.  This adds additional support to our choice of adopting the $L_{\rm bol}$ value measured directly on the \textit{Swift} SED for our analysis.

It is important to emphasize that the bolometric correction relations of \citep{lusso2010A&A...512A..34L} and \citep{Duras2020A&A...636A..73D} are carefully calibrated using large samples of X-ray selected, persistently bright AGN. While based on a statistically large database of high-quality, carefully selected data the method caries two characteristic traits to be aware of. Firstly, the nature of determining an average bolometric correction factor (eqn.~\ref{eq:duras}) or a prescription based on a fit to data with object-to-object scatter (eqn.~\ref{eq:lusso}) inherently assumes that the intrinsic SED of the object is very similar to that of the population average. Secondly, the population average SED and prescription will depend on the sample selection.  The methods by \citep{lusso2010A&A...512A..34L} and \citep{Duras2020A&A...636A..73D} are based on X-ray bright, commonly observed Type 1 AGN that are generally not highly variable, as observed for CLAGN.

In contrast, as demonstrated in this work, Mkn~590 likely has a non-standard SED (Fig.~\ref{fig:compare}) and, on top, Mkn~590 has undergone dramatic changes in both X-ray and optical/UV luminosity, reflecting substantial restructuring of its inner accretion flow. These transitions can potentially alter the thermal balance between disk and corona emission on timescales much shorter than the averaging timescales used in population studies \citep{mcleod2016MNRAS.457..389M,2018Ross}. Consequently, bolometric corrections derived from narrow band luminosities may not accurately capture the instantaneous accretion power of a source undergoing rapid state changes.

Furthermore, recent variability studies further show that CLAGNs exhibit flux evolution patterns that are distinct from typical Seyfert 1/2 AGN \citep{rambaugh2018ApJ...854..160R,wang2025arXiv251110217W}, reinforcing the need for source specific estimates of $L_{\rm bol}$. They provide a more physically representative estimate of the $L_{\rm bol}$ for CLAGNs, where disk-corona coupling evolves dynamically and cannot be reliably captured by static bolometric correction factors. 
\begin{figure}[ht]
    \centering
    \includegraphics[scale=.70]{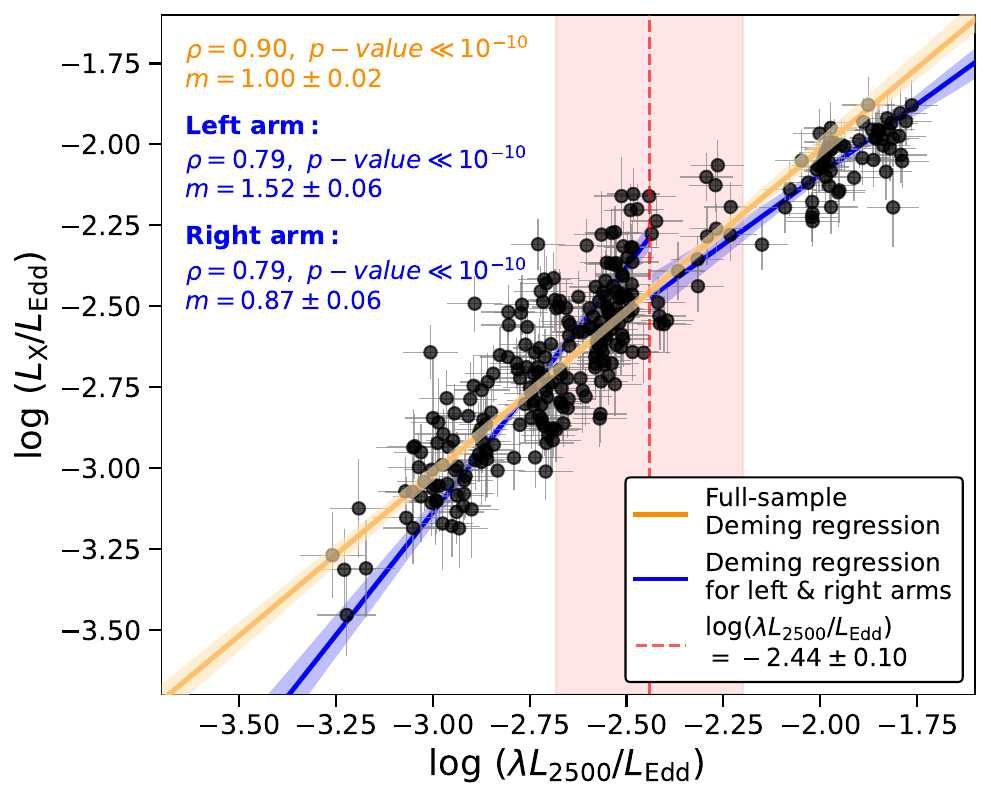}
    \caption{Comparison of the UV ($\lambda L_{2500}/L_{\rm Edd}$) and X-ray-based ($ L_{X}/L_{\rm Edd}$) Eddington ratios described in Sec.~\ref{sec:alfy}. The red dashed line marks the best-fitting breakpoint obtained from the piecewise linear regression, while the shaded region denotes its 95\% confidence interval. The solid orange line shows the Deming regression fitted to the full dataset, whereas the solid blue lines show independent Deming fits to the data below and above the breakpoint. The corresponding best-fitting slopes ($m$) are listed in the figure. Spearman correlation coefficients ($\rho$) are also shown, demonstrating a strong correlation for both the full sample and the two individual branches.}
    \label{fig:LX_uv}
\end{figure}
In conclusion, for the reasons outlined above, we compute $L_{\rm bol}$ by directly integrating the intrinsic broadband SED for each simultaneous {\it Swift}-XRT and UVOT measurements, and use it for all scientific interpretations in this work.

To examine whether the UV- and X-ray-based Eddington-ratio tracers themselves exhibit any change in their mutual scaling, we applied the same piecewise linear regression analysis used throughout this work to the $\log (\lambda L_{2500}/L_{\rm Edd})$ vs $\log (L_{X}/L_{\rm Edd})$ relation. We identify a single break point at $ \log (\lambda L_{2500}/L_{\rm Edd})$ = -2.44$\pm$0.10 with a null-hypothesis probability of p $\ll$ 10$^{-10}$, indicating a statistically significant departure from a single linear relation. We then fitted the entire dataset with a single Deming regression, as well as separately on either side of the inferred break. The resulting best-fit relations are shown in Fig.~\ref{fig:LX_uv}, together with the corresponding Spearman rank coefficients. Clearly, the two Eddington-ratio tracers are strongly correlated during the monitoring campaign, potentially suggesting that they trace the same underlying changes in accretion power. The single Deming regression (solid orange line) provides a good first-order description of the relation, with small deviation from linearity. However, the individual Deming regression on each branch (solid blue lines) reveal a steeper slope below the break and a shallower slope above it.

\section{Statistical tools used in this paper}
\label{appendixD}

To determine the turnover of the $\aox-\eddfrac$ relation we use \texttt{Piecewise linear regression}, also known as break point analysis or segmented regression. The tool is useful for investigating data that exhibit one or more changes in gradient. The \texttt{piecewise-regression} Python package implements this methodology by fitting continuous piecewise linear models in which both the break point locations and the segment slopes are estimated simultaneously. The approach follows the iterative algorithm of \cite{muggeo,muggeo2}, in which the non-linear model is linearized about initial break point estimates and refitted using ordinary least squares regression until convergence. The package further provides a comprehensive statistical framework, including confidence intervals for all model parameters and formal hypothesis testing for the presence of breakpoints. It incorporates extensive bootstrapping across many random realizations to ensure convergence toward the global minimum solution \citep{bootstrap}. Finally, it uses the Davies test to estimate the probability of existence of at least one break point against the  null hypothesis of no break point(s) \citep{davies}.

We also implemented the \texttt{Deming regression} to fit linear models to either side of the break in $\aox-\eddfrac$ relation. 
The \texttt{Deming regression} is an errors-in-variables regression method that accounts for measurement uncertainties in both the dependent and independent variables \citep{deming}. It works by fitting a straight line and minimizing the squared orthogonal distances between the data points and the model, rather than the vertical residuals, while explicitly accounting for measurement uncertainties in both variables through their variance ratio. To estimate the uncertainty in fitted slope and intercept and determine the 95\% confidence interval, we generated 5000 bootstrap re-samplings of the dataset and refitted each realization, from which we derived the corresponding 95\% confidence intervals.

\bibliography{sample701}{}
\bibliographystyle{aasjournalv7}



\end{document}